\documentclass[onecolumn,12pt]{article}

\usepackage{graphicx} 
\usepackage{setspace} 
\usepackage{graphics,epsfig}
\usepackage{amsmath}
\usepackage{amssymb}
\usepackage{bm}
\usepackage{xcolor}
\usepackage{multirow}

\usepackage{subfigure}
\usepackage{txfonts}
\usepackage[a-1b]{pdfx}
\usepackage{hyperref}
\usepackage{float}

\newcommand*{\obs}{\ensuremath{\mathbf{Y}}}

\newcommand{\btheta}{\bm{\theta}}
\newcommand{\bSigma}{\bm{\Sigma}}

\UseRawInputEncoding

 \usepackage[top=2cm, bottom=4cm, left=2.25cm, right=2.25cm]{geometry}

\begin{document}

\title{Adaptive posterior distributions for uncertainty analysis of covariance matrices in Bayesian inversion problems for multioutput signals}
\author{E. Curbelo$^{*}$, L. Martino$^{\dagger}$, F. Llorente$^{**}$,  D. Delgado-Gomez$^{*}$, \\  
{\small  $^{*}$ Universidad Carlos III de Madrid (UC3M),  Madrid, Spain.}\\
{\small $^{\dagger}$ Universit{\`a} di Catania, Catania, Italia. } \\
{\small  $^{**}$ Stony Brook University, New York, USA.} \\
}

\date{ }
\maketitle

\begin{abstract}
	In this paper we address the problem of performing Bayesian inference for the parameters of a nonlinear multioutput model and the covariance matrix of the different output signals.   
We propose an adaptive importance sampling (AIS) scheme for multivariate Bayesian inversion problems, which is based in two main ideas: the variables of interest are split in two blocks and the inference takes advantage of known analytical optimization formulas.
We estimate both the unknown parameters of the multivariate non-linear model and the covariance matrix of the noise. In the first part of the proposed inference scheme, a novel AIS technique called adaptive target adaptive importance sampling (ATAIS) is designed, which alternates iteratively between an IS technique over the  parameters of the non-linear model and a frequentist approach for the covariance matrix of the noise. In the second part of the proposed inference scheme, a prior density over the covariance matrix is considered and the cloud of samples obtained by ATAIS are recycled and re-weighted  to obtain a complete Bayesian study over the model parameters and covariance matrix. ATAIS is the main contribution of the work. Additionally, the inverted layered importance sampling (ILIS) is presented as a possible compelling algorithm (but based on a conceptually simpler idea). Different numerical examples show the benefits of the proposed approaches. 
\newline
{\bf Keywords:} Bayesian inversion; importance sampling; uncertainty analysis; covariance matrix; tempering;  sequence of posteriors
\end{abstract}

\maketitle

\section{Introduction}

The estimation of parameters from noisy observations is at the center of areas such as signal processing, statistics and machine learning. Looking at this problem from a Bayesian perspective, the inference problem becomes the construction and analysis of the posterior density over the unknown parameters \cite{Fitzgerald01,Andrieu:MCMachineLearning2003}. The computation of complicated integrals involving these posterior distributions is often needed (e.g., any moment of the random variable distributed as the posterior density). Monte Carlo sampling methods are able to draw samples from the posterior probability density function (pdf) and hence those integrals can be approximated by stochastic quadrature formulas employing the generated samples. The Monte Carlo techniques can be divided into four main families: direct transformation methods, rejection sampling, importance sampling and Markov Chain Monte Carlo (MCMC) algorithms \cite{Robert04,7974876,LuengoSurvey,MARTINO2018134}. The last two classes are the most used by the users, since they are universal methods, i.e., they can always be applied.
\newline
\newline
However, the Monte Carlo techniques find several difficulties that jeopardize their performance  in many scenarios, for instance, when working in high-dimensional spaces, and with  narrow, tight posteriors. Both issues are related to the problem of the exhaustive exploration of the state space. For these reasons, many Monte Carlo algorithms try to work in sub-dimensional spaces (step by step, with iterative or sequential procedures), such as the Gibbs sampling and the particle filtering schemes \cite{LuengoSurvey,RecGibbs,Doucet08tut,Djuric03}. { Furthermore, the inference on a positive definite matrix requires to fulfill certain restrictions about the entries of the matrix. This issue makes also difficult the use of gradient approaches since many solutions can be outside the allowed support domain. }
\newline
\newline
In this work, we focus on the problem of making a  joint inference on a covariance matrix and a vector of parameters \cite{Robert04,Liu04}. This is a particularly complex inference problem
since bad choices of the covariance matrix can jeopardize the sampling of the vector of interest  \cite{AdaptiveSequentialEstimationWithUnknownNoiseStatistics, BayesianEstimationInMultivariateInterLaboratoryStudiesWithUnknownCovarianceMatrices, BayesianInferenceOfAMultivariateRegressionModel}. This problem can suffer both issues previously  described: it is often high - dimensional (especially if the dimension matrix is big) and the posterior is often tight.
More specifically,  we address a generic multidimensional Bayesian inversion problem, where each vector observation $\bf{y}_r$ is the output of a multidimensional, nonlinear {\it vectorial} mapping $\bf{f}(\bm{\theta})$ of the parameter of interest  $\bm{\theta}$, perturbed by an error vector with correlated components that, e.g., can be Gaussian ${\bf v}_r \sim \mathcal{N}({\bf v}_r|\bm{0}, \bm{\Sigma})$.\footnote{We assume Gaussianity in the first part of the work, only for clarity and simplicity in the explanation.} The goal is to make inference in the joint space of $\bm{\theta}$ and $\bm{\Sigma}$. The dimension of the entire space grows linearly with the dimension of the vector $\bm{\theta}$ and quadratically  with the dimension of the matrix $ \bm{\Sigma}$. 
We consider virtually no assumptions over the vectorial non-linearity $\bf{f}$, and usually it represents some complex physical process.  For instance, $\bf{f}(\bm{\theta})$ could also be non-differentiable. In this work, the unique requirement about $\bf{f}$ is to be able to evaluate point-wise  $\bf{f}(\bm{\theta})$. Since, the inference task on the complete space $\{\bm{\theta}, {\bm \Sigma}\}$ is particularly challenging, we introduce two different compelling Monte Carlo schemes based on the idea of splitting the inference space into two blocks, $\bm{\theta}$ and $ {\bm\Sigma}$ (as in a block Gibbs sampling \cite{Robert04,RecGibbs}, and/or other similar approaches \cite{rothman2010sparse}).  
\newline
\newline
{\bf Main proposed scheme - ATAIS.} Firstly, we extend and generalize the approach presented in \cite{ATAIS,ATAIS2}. The proposed inference scheme is divided into two main parts. 
 In the first part, we approximate the conditional posterior of $\bm{\theta}$ given the data and  the maximum likelihood estimator  $ {\bm\Sigma}_{\texttt{ML}}$ of the matrix ${\bm \Sigma}$. This first part is called {\it adaptive target adaptive importance sampling} (ATAIS), since we perform an adaptive importance sampling on a sequence of adaptive posteriors (due to the variation of $ {\bm\Sigma}$).  The ATAIS method is then completed by a second part which allows a complete Bayesian inference also over $\bm{\Sigma}$. Indeed, in this second inference part, we approximate the complete posterior of pair of variables of interest $\{\bm{\theta}, {\bm\Sigma}\}$,  without any additional generation of samples over $\bm{\theta}$. The resulting scheme is a robust inference approach for Bayesian inversion, based on an adaptive importance sampler that addresses a sequence of different conditional posteriors and a post-process that allows a Bayesian inference over $ {\bm\Sigma}$ as well.  
 We refer to the overall scheme (first and second part) as {\it complete ATAIS}. The conditional posteriors addressed by ATAIS differ in the use of different covariance matrices: this procedure can resemble a tempering of the posterior distribution \cite{kirkpatrick83,Marinari92,Friel08,Moral06}.
  \newline
\newline
{\bf Auxiliary competitive scheme - ILIS.} As also remarked in different works \cite{ATAIS,FastBayesianInversionForHighDimensionalInverseProblems, AdaptiveSequentialEstimationWithUnknownNoiseStatistics}, the application of a Monte Carlo sampling method directly in  the complete space $\{\bm{\theta}, {\bm \Sigma}\}$ is particularly challenging and the resulting performance is quite poor. Hence, at least with our current knowledge of the literature, it is also difficult to find a competitive alternative to ATAIS, which can provide errors in estimation of the same magnitude. However, in our practical experience, we have designed another Monte Carlo scheme (conceptually simpler than ATAIS) that can also obtain reasonable results. We call this competitive scheme, {\it inverse layered importance sampling} (ILIS) since we adapt the idea given  \cite{LAIS17, Llorente22mcmcDriven} for this inference context. With respect to the main algorithm in  \cite{LAIS17, Llorente22mcmcDriven}, we switch the positions of the importance sampling (IS) method and Markov Chain Monte Carlo (MCMC) techniques  \cite{Cornuet12,Bugallo15,RecGibbs,Liu04}: in ILIS the upper layer is formed by an IS procedure, and the lower layer is formed by {\it weighted} MCMC chains. Conceptually speaking, ILIS can be considered simpler than ATAIS, but the ILIS performance is more sensible on choice of certain proposal parameters (e.g., covariance reference matrix in the upper layer proposal), whereas the complete ATAIS procedure is able to auto-tune some auxiliary parameters, reducing the number of parameters decided by the user. In this sense, ATAIS is more automatic and robust than ILIS. As a final observation, we highlight that ATAIS could also be  combined and jointly employed. 
\newline
\newline
A summary of the main contributions of the work and related important considerations are given below:
\begin{itemize}
\item {We propose a robust and efficient inference scheme for complex Bayesian inversion problems, where a scale (covariance) matrix must also be  estimated. The model considers a vectorial non-linear function ${\bf f}(\bm{\theta})$ to invert, that can represent complex dynamical systems, a set of time series  models, or a statistical spatial model for instance.}
\item { The proposed method allows a complete Bayesian analysis of   $\bm{\theta}$ and ${\bm\Sigma}$ so that, we can perform uncertainty analysis over $\bm{\theta}$ and/or ${\bm\Sigma}$, obtaining credible intervals. Moreover, we can perform hypothesis testing or model selection approximating the marginal likelihood {\color{black}\cite{Robert04, Llorente20, Llorente22safePriors}}. Hence, we remark that the proposed scheme is {\it much more} than an optimizer: it is a sampler that allows a complete Bayesian inference over  $\bm{\theta}$ and ${\bm\Sigma}$. In its second inference part, ATAIS recycles all the samples (w.r.t. ${\bm \theta}$) and the posterior evaluations from the first part. Thus, this second part  does not require any additional evaluation of the possibly complex and costly nonlinearity ${\bf f}(\bm{\theta})$.}
\item {ATAIS can be considered as an adaptive importance sampler where {\it both}  proposal and target pdfs are adapted. Indeed, in ATAIS, we consider a sequence of adaptive conditional posteriors.  The complete ATAIS method can be also interpreted as an IS version of 
 the {\it recycling Gibbs sampling} scheme in \cite{RecGibbs} (with two blocks). Indeed, the complete space is divided into two blocks, $\bm{\theta}$ and ${\bm\Sigma}$, where different numbers of samples are considered for each block, denoted as $NT$ for ${\bm \theta}$ and $J$ for  ${\bm \Sigma}$.}
 \item {We also introduce several extensions as addressing models with non-Gaussian noises, e.g., with $t$-Student's noise (or, more generally, with other elliptical distributions) and/or the possible use of mini-batches (that is allowed by ATAIS). A discussion with practical suggestions regarding the tuning of hyper-parameters of the prior densities is provided. Furthermore, a detailed discussion about alternatives and improvements (for speeding up or reducing the computational cost) is given as well.  } 
\item We also designed  a competitive sampling scheme, denoted as ILIS, for comparing the performance of ATAIS.  Several comparisons with other benchmark techniques are also provided.
\end{itemize}
The range of applications includes the inference of any temporal and spatial dataset such as: systems of differential equations explaining the behavior of a disease (for instance, SIR
models for COVID) or time series in different cities, weather prediction in different regions, medical signals with multiple sensors, topology graph estimation and, more generally, any inference problem with time-varying signals defined in different spatial locations \cite{huang2020bayesian, beira2021differential, reaz2006techniques,qu2024joint}.
\newline
The paper is structured as follows. We start with the description of the problem statement in Section \ref{sec:bayes}. The first part of the main proposed inference scheme is introduced in Sections \ref{FirstPartATAIS} and \ref{ATAIS_sect}. {Section \ref{SectComp} provides a detailed discussion about possible alternatives and improvements.} 
 The second part of the main proposed inference scheme is described in Section \ref{SectSuperFer}.
The alternative scheme, inverted layered importance sampling (ILIS),  is given in Section \ref{ILISsect}. Finally, Section \ref{sec:Simul} contains several numerical experiments and Section \ref{conclSect} provides some final conclusions.

\section{Problem Statement}
\label{sec:bayes}

%
Let us denote as ${\bm \theta}=[ \theta_1,..., \theta_M]^{\top} \in {\bm \Theta} \subseteq \mathbb{R}^{M}$, a variable of interest that we desire to infer.
Moreover, related to ${\bm \theta}$, we observe
\begin{itemize}
\item  $R$ values in different time instants (or spatial points) of 
 \item $K$ different signals (time series), i.e., 
\end{itemize}
  $\mathbf{y}_r=[y_{r,1},..., y_{r,K}]\in  \mathbb{R}^{K\times 1}$ for $r=1,...,R$. Hence, all received data can be stored in a matrix ${\bf Y}=[\mathbf{y}_1,...,\mathbf{y}_R]\in \mathbb{R}^{K\times R}$.
 Furthermore, let  us consider the observation model 
\begin{align}\label{EqObsModel_all}
{\bf y}_r&={\bf f}_r({\bm \theta})+\mathbf{v}_r,   \qquad r=1,...,R,
\\
\mathbf{Y}&= {\bf F}(\boldsymbol{\theta}) + \mathbf{V}, 
\label{eq:model}
\end{align}
where we have a nonlinear mapping for each time instant and each time series,
\begin{align}
{\bf f}_r({\bm \theta})&=[f_{r,1}({\bm \theta}),..., f_{r,K}({\bm \theta})]^{\top}:  {\bm \Theta} \subseteq \mathbb{R}^{M} \rightarrow   \mathbb{R}^{K\times 1}, \\
{\bf F}(\boldsymbol{\theta})&=[{\bf f}_1({\bm \theta}),...,{\bf f}_R({\bm \theta})]: {\bm \Theta} \subseteq \mathbb{R}^{M} \rightarrow   \mathbb{R}^{K\times R}, 
\end{align}
and a $K\times 1$ vector of Gaussian noise perturbation for each time instant,
\begin{align}
\mathbf{v}_r&=[v_{r,1},...,v_{r,K}]^{\top}\sim \mathcal{N}({\bf v}_r|{\bf 0}, {\bm \Sigma} ) \in \mathbb{R}^{K\times 1}, \label{EqGaussNoise}\\
\mathbf{V}&=[{\bf v}_{1},...,{\bf v}_{R}] \in \mathbb{R}^{K\times R},
\end{align}
 where ${\bm \Sigma}$ is  $K\times K$  covariance matrix, which generally is unknown.  The mapping ${\bf f}_r(\boldsymbol{\theta})$ could be analytically unknown,  the only assumption is that we are able to evaluate it pointwise.\footnote{Each component $f_{r,k}$, for $k=1,...K$, can be a function of the complete vector ${\bm \theta}$ or only a subset of components of this vector. See for instance the simulation experiment in Section \ref{SectMO}.} The likelihood  function is 
\begin{eqnarray}
 \ell({\bf Y}|\boldsymbol{\theta}, {\bm \Sigma})&=&
 \left(\frac{1}{(2\pi)^{K/2} \mbox{det}({\bm \Sigma})^{1/2}}\right)^R \exp\left( -\frac{1}{2} 
 \left[ 
\sum_{r=1}^R\left(\mathbf{y}_r - {\bf f}_r(\boldsymbol{\theta})\right)^{\top}{\bm \Sigma}^{-1} \left(\mathbf{y}_r - {\bf f}_r(\boldsymbol{\theta})\right)
  \right] 
 \right), 
\label{eq:LH}
\end{eqnarray}
Note that we have two types of variables of interest for an inference point of view:
\begin{itemize}
\item the vector $\boldsymbol{\theta}$ contains the parameters of the nonlinear mapping ${\bf f}_r(\boldsymbol{\theta})$, for $r=1,...,R$,
\item and ${\bm \Sigma}$ is a scale matrix of the likelihood function.
\end{itemize}
Given the complete matrix of measurements ${\bf Y}$, we desire to make inferences regarding the hidden parameters ${\bm \theta}$ and the noise matrix ${\bm \Sigma}$, obtaining at least some point estimators ${\widehat {\bm \theta}}$ and ${\widehat {\bm \Sigma}}$. 
We are also interested in performing uncertainty and correlation analysis among the components of $\bm{\theta}$.
Furthermore, we aim to perform model selection, i.e., to compare, select  or properly average different models.

\subsection{Application to time series and spatial processes} 

\begin{figure}[h!]
	\centering
	\includegraphics[scale=0.5]{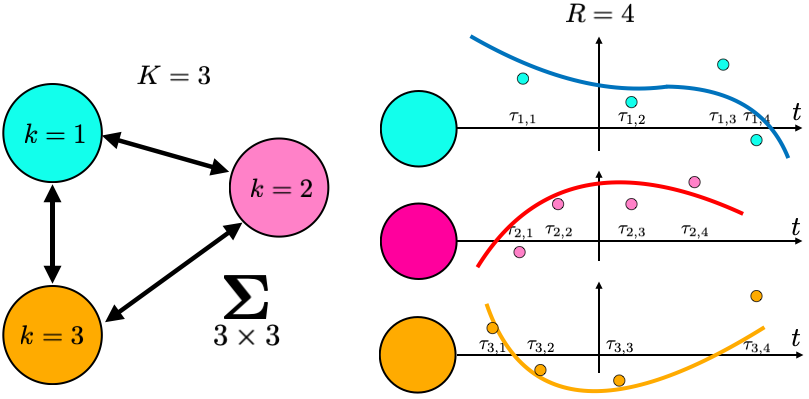}
	\caption{Graphical representation of the considered multioutput model with $K=3$ output signals and $R=4$ time instants for each signal. One can suppose that ${\bm \Sigma}$ represents a  $3\times  3$ covariance matrix of three possible nodes in a graph.} \label{SuperFig}
\end{figure}

The range of applications of the considered model is very broad. For instance, in the case of having $K$ different time series (in continuous or discrete time), or $K$ spatial processes we can have 
more explicit notation, where there is a one-to-one correspondence between each index $r\in \{1,...,R\}$ and a real time instant $\tau_{k,r} \in \mathbb{R}$ or a point ${\bf x}_{k,r}\in \mathbb{R}^{d\times 1}$, i.e.,
$$
r\in \{1,...,R\} \longleftrightarrow \{\tau_{k,r} \in \mathbb{R}\}_{k=1}^K , \qquad r \in \{1,...,R\}\longleftrightarrow \{{\bf x}_{k,r}\in \mathbb{R}^d\}_{k=1}^K.
$$
Each vector ${\bf y}_r$ (of dimension $K\times 1$) contains the measurements at time instants $\tau_{1,r},...,\tau_{K,r}$ (or ${\bf x}_{1,r},...,{\bf x}_{K,r}$) each one corresponding to a different time series. {Generally, an alternative notation is
 \begin{align}\label{NotationAqui}
{\bf y}={\bf f}({\bm \theta},{\bm \tau})+\mathbf{v}, \qquad {\bf y}={\bf f}({\bm \theta},{\bf X})+\mathbf{v}, 
\end{align}
where ${\bm \tau}=[\tau_{1},...,\tau_{K}]$ and  ${\bf X}=[{\bf x}_{1},...,{\bf x}_{K}]$ is $d\times K$ matrix.\footnote{{ Note that in the isotopic scenario, where $\tau_{1}= \tau_{2}...=\tau_{K}=\tau$, the notation can be simplified  as ${\bf f}({\bm \theta}, \tau)$ where $\tau$ is a scalar.}}} More specifically, recalling the observation equation 
${\bf y}_r={\bf f}_r({\bm \theta})+\mathbf{v}_r$,
we could use a more explicit notation, instead of ${\bf f}_r({\bm \theta})$, i.e.,
\begin{align}
{\bf y}_r&={\bf f}({\bm \theta},\tau_{1,r},...,\tau_{K,r})+\mathbf{v}_r,  \\
{\bf y}_r&={\bf f}({\bm \theta},{\bf x}_{1,r},...,{\bf x}_{K,r})+\mathbf{v}_r,  \qquad r=1,...,R,
\label{eq:model_otro}
\end{align}
where $\tau_{1,r}$, ..., $\tau_{K,r}$, or ${\bf x}_{1,r}$, ..., ${\bf x}_{K,r}$ play the role of auxiliary known parameters (or vectors of parameters). A graphical representation is given in Figure \ref{SuperFig}.

\noindent
{\bf Remark.} Note that the vector  ${\bm \theta}=[ \theta_1,..., \theta_M]^{\top}$ contains the parameters that are possibly shared from the models representing the $K$ different time series (or $K$ spatial processes), or all the parameters that only affect one series (or just a subset of time series).

\subsection{Bayesian inference in the complete space} 

The full Bayesian solution considers the study of 
the complete posterior density
\begin{eqnarray}
\label{Total_post}
p({\bm \theta}, {\bm \Sigma}|{\bf Y})= \frac{1}{p({\bf Y})} p({\bm \theta}, {\bm \Sigma},{\bf Y})= \frac{1}{p({\bf Y})}\ell({\bf Y}|\boldsymbol{\theta}, {\bm \Sigma}) g_\theta({\bm \theta})g_{\bm \Sigma}({\bm \Sigma}), 
\end{eqnarray}
 where $g_\theta({\bm \theta})$ and $g_{\bm \Sigma}({\bm \Sigma})$ represent the prior densities over the vector ${\bm \theta}$ and the matrix ${\bm \Sigma}$.  Usually, complex integrals involving $p({\bm \theta}, {\bm \Sigma}|{\bf Y})$ should be computed in order to perform the inference.   
\newline
\newline
\textbf{Main observation.} Generally, generating random samples from a complicated posterior in Eq. \eqref{Total_post} and computing efficiently the integrals involving $p({\bm \theta}, {\bm \Sigma}|{\bf Y})$ is a hard task. Note that the complete dimension of the inference problem $D$ is 
  $$
D=  M + \dfrac{K(K+1)}{2},
 $$ 
 i.e., the number of parameters to infer is exactly $D$. With $M=2$ and $K=5$, we have $D=17$ and with $M=2$, $K=10$ we have $D=57$. The dimension $D$ grows linearly  with $M$ and quadratic with respect to $K$. Moreover, we also have  the constraints regarding  ${\bm \Sigma}$, since it must be a covariance matrix. This task becomes even more difficult when we try to perform a joint inference, learning jointly the covariance matrix ${\bm \Sigma}$ and  parameters of the nonlinearity  ${\bm \theta}$. Indeed, ``wrong choices'' of ${\bm \Sigma}$ can easily jeopardize the sampling of ${\bm \theta}$.  { Note also that the inference of a covariance matrix requires to fulfill certain constraints on the entries of the matrix. }
\newline 
\newline
\textbf{Proposed approach.} Below, we describe an inference scheme formed by {\it two main parts}. First, we tackle the problem of drawing from the conditional posterior of ${\bm \theta}$,  conditioned to the data and the maximum likelihood (ML) estimator of ${\bm \Sigma}$ (that is generally unknown and must be approximated as well). 
Therefore, in this first part, we apply a Bayesian inference over ${\bm \theta}$ and a frequentist approach over ${\bm \Sigma}$ (obtaining an approximation of the ML estimator). In the second part, we assume also a prior density over the covariance matrix ${\bm \Sigma}$, and perform a Bayesian inference over ${\bm \Sigma}$ as well, recycling the outputs (weighted samples and other information) obtained in the first part. 


\section{First part of the proposed inference scheme}\label{FirstPartATAIS}

The main idea underlying the proposed inference scheme is to take advantage of the split of the inference space (working firstly in smaller portions of the entire space). In the first part, described in this section, we search for high probability regions in the complete space, sampling from a {\it sequence of adaptive conditional posterior distributions} with respect to ${\bm \theta}$ (given a covariance matrix ${\bm \Sigma}_{\texttt{ML}}$). An analytic formula is employed for obtaining a sequence of optimized matrices ${\bm \Sigma}_{\texttt{ML}}$. In the second part, described in Section \ref{SectSuperFer}, we generate random matrices from a {\it tuned} prior pdf  (or possibly other proposal density) and we re-weight all the previously generated samples w.r.t. ${\bm \theta}$,  in order to allow a complete Bayesian inference (hence including uncertainty analysis, etc.) for both ${\bm \theta}$ and ${\bm \Sigma}$.
\newline
\newline
More specifically, in the first stage, we consider a sub-optimal (in Bayesian sense) but substantially more efficient inference scheme (since we work in a reduced - much smaller - dimensional space), studying only a sequence of conditional posterior distributions. More precisely, we study the following conditional posterior
\begin{align}
p({\bm \theta}|{\bf Y}, {\bm \Sigma}_{\texttt{ML}}) &= \frac{\ell({\bf Y}|{\bm \Sigma}_{\texttt{ML}},{\bm \theta}) g_\theta({\bm \theta})}{p({\bf Y}|{\bm \Sigma}_{\texttt{ML}})}  \propto \ell({\bf Y}|{\bm \Sigma}_{\texttt{ML}},{\bm \theta}) g_\theta({\bm \theta}). \label{EstaEslaConDMargPost} 
\end{align}   
Furthermore, we have denoted the {\it conditioned} maximum likelihood estimator of ${\bm \Sigma}$  as
\begin{align}
{\bm \Sigma}_{\texttt{ML}}=\arg \max_{{\bm \Sigma}} \ell({\bf Y}|{\bm \Sigma},{\bm \theta}_\texttt{MAP}),
\end{align} 
where $\bm{\theta}_\texttt{MAP}$ denotes the global maximum of $p(\bm{\theta}|\obs, {\bm \Sigma}_{\texttt{ML}})$, i.e.,
\begin{align}
	\bm{\theta}_\texttt{MAP} 
	 &=\arg\max_{\bm{\theta}}  \log p(\bm{\theta}|\obs, {\bm \Sigma}_{\texttt{ML}}), \nonumber \\
	 &=\arg\min_{\bm{\theta}}  \left[ \sum_{r=1}^R\left(\mathbf{y}_r - {\bf f}_r(\boldsymbol{\theta})\right)^{\top}{\bm \Sigma}_{\texttt{ML}}^{-1} \left(\mathbf{y}_r - {\bf f}_r(\boldsymbol{\theta})\right) + \log g_{\theta}({\bm \theta})\right].
\end{align}
It is important to observe that, given $\bm{\theta}_\texttt{MAP}$, we have the analytic form of ${\bm \Sigma}_{\texttt{ML}}$, i.e.,
\begin{align}\label{MatrixTRUEeq}
{\bm \Sigma}_{\texttt{ML}}=\frac{1}{R} \sum_{r=1}^R\left({\bf y}_r- {\bf f}_r({{\bm \theta}}_{\texttt{MAP}})\right)\left({\bf y}_r- {\bf f}_r({{\bm \theta}}_{\texttt{MAP}})\right)^{\top}.
\end{align}
Note that ${\bm \Sigma}_{\texttt{ML}}$ depends on $\bm{\theta}_\texttt{MAP}$, and $\bm{\theta}_\texttt{MAP} $ depends on ${\bm \Sigma}_{\texttt{ML}}$. { For high dimensional estimators of ${\bm \Sigma}_{\texttt{ML}}$, see the procedure described in \cite{ledoit2004well}. Moreover, in case of heavy-tailed distributions and presence of outliers, see the relevant survey \cite{ke2019user}.}  Finally, similar approaches for dealing with unknown covariance can be found in \cite{FastBayesianInversionForHighDimensionalInverseProblems, AdaptiveSequentialEstimationWithUnknownNoiseStatistics}.
\newline
\newline
{\bf Remark.} The key idea to implement this inference scheme is to perform an alternating optimization procedure where, at each iteration $t$, we produce two estimations $\widehat{\bm{\theta}}_\texttt{MAP}^{(t)}$, $\widehat{{\bm \Sigma}}_{\texttt{ML}}^{(t)}$ of $\bm{\theta}_\texttt{MAP} $, ${\bm \Sigma}_{\texttt{ML}}$, respectively \cite{BayesianEstimationInMultivariateInterLaboratoryStudiesWithUnknownCovarianceMatrices, BayesianInferenceOfAMultivariateRegressionModel}. Clearly, we desire the convergence  as the number of iterations grows, $t \rightarrow \infty$, i.e.,
\begin{align}
\widehat{\bm{\theta}}_\texttt{MAP}^{(t)}& \longrightarrow {\bm \theta}_\texttt{MAP}, \\
\widehat{\bm{\Sigma}}_\texttt{ML}^{(t)}& \longrightarrow {\bm \Sigma}_\texttt{ML}. 
\end{align} 
The suggested iterative approach is summarized briefly in two steps. Starting with an initial matrix  $\mathbf{\Sigma}_{\texttt{ML}}^{(0)}$, that is as a rough approximation of $\widehat{\mathbf{\Sigma}}_{\texttt{ML}}$, the alternating optimization procedure is given in Table \ref{TableAO}. 
\begin{table}[!h]
	\centering
	\caption{Alternating optimization.} \label{TableAO}
\vspace{-0.3cm}
\begin{center}
\begin{tabular}{|p{0.95\columnwidth}|} 
 \hline
For $t=1,\dots,T$: 
\begin{itemize}
	\item[1]  Estimate, by Monte Carlo,
	\begin{align}\label{EqMAPAqui}
	\bm{\theta}_\texttt{MAP}^{(t)}  =\arg\min_{\bm{\theta}}  \left[ \sum_{r=1}^R\left(\mathbf{y}_r - {\bf f}_r(\boldsymbol{\theta})\right)^{\top}\left[\widehat{{\bm \Sigma}}_{\texttt{ML}}^{(t-1)}\right]^{-1} \left(\mathbf{y}_r - {\bf f}_r(\boldsymbol{\theta})\right) - \log g_{\theta}({\bm \theta})\right],
\end{align}
	obtaining $\widehat{\bm{\theta}}_\texttt{MAP}^{(t)}$, e.g., using an importance sampling (IS) scheme with respect to $p(\bm{\theta}|\obs,\widehat{{\bm \Sigma}}_\texttt{ML}^{(t-1)})$.
	
	\item[2] Compute 
	\begin{align}\label{EqSigmaAqui}
\widehat{{\bm \Sigma}}_{\texttt{ML}}^{(t)}=\frac{1}{R} \sum_{r=1}^R\left({\bf y}_r- {\bf f}_r({\widehat{\bm \theta}}^{(t)}_{\texttt{MAP}})\right)\left({\bf y}_r- {\bf f}_r({\widehat{\bm \theta}}^{(t)}_{\texttt{MAP}})\right)^{\top}.
\end{align}



\end{itemize}  \\
 \hline
\end{tabular}
\end{center}
\end{table}

\noindent
Since, we employ IS scheme for obtaining $\widehat{\bm{\theta}}_\texttt{MAP}^{(t)}$, at each $t$-th iteration, we have also a cloud of particles $\{{\bm \theta}_t^{(n)})\}_{n=1}^N$ that can be used for performing Bayesian inference over $\bm{\theta}$.
 Namely, after $T$ iterations, we can build a particle approximation of $p(\bm{\theta}|\obs,\widehat{{\bm \Sigma}}_\texttt{ML}^{(T)})$, i.e., 
\begin{align}\label{FinalParApprox}
\widehat{p}(\bm{\theta}|\obs,\widehat{{\bm \Sigma}}_\texttt{ML}^{(T)})=\sum_{t=1}^T \sum_{n=1}^N \widetilde{w}_{t}^{(n)} \delta({\bm \theta}-{\bm \theta}_t^{(n)}), \qquad\quad \displaystyle\sum_{t=1}^T\sum_{n=1}^N \widetilde{w}_{t}^{(n)}=1.
\end{align}
 By Eq. \eqref{FinalParApprox}, we can approximate all the moments associated with the conditional posterior $p(\bm{\theta}|\obs,\widehat{{\bm \Sigma}}_\texttt{ML}^{(T)})$ hence, for instance, we can also provide an uncertainty estimation over the vector of $\bm{\theta}$.
 \newline
\newline
{\bf On the convergence of the alternating optimization.} Due to the error in step 1 of the alternating optimization (described above) can be controlled by the number of particles $N$ (i.e., the error in the approximation of $\bm{\theta}_\texttt{MAP}$ can be bounded by increasing $N$, i.e., even with a bad choice of $\widehat{\bm{\Sigma}}_{\texttt{ML}}^{(t-1)}$ we can obtain a reasonable vector $\widehat{\bm{\theta}}_\texttt{MAP}^{(t)}$, and the estimator $\widehat{\bm{\Sigma}}_{\texttt{ML}}^{(t)}$ in Eq. \eqref{EqSigmaAqui} approaches the matrix ${\bm \Sigma}_{\texttt{ML}}$ in Eq. \eqref{MatrixTRUEeq}, as $t\rightarrow \infty$.
 Moreover, as the number of realizations $R$ grows the matrix ${\bm \Sigma}_{\texttt{ML}}$ in Eq. \eqref{MatrixTRUEeq} converges to the true covariance matrix of the data. 
 \newline
 Note that the pair $\bm{\theta}_\texttt{MAP}$ and  ${\bm \Sigma}_{\texttt{ML}}$  are {\it fixed points} of {\it the iterative (dynamical) system} formed by Eqs. \eqref{EqMAPAqui}-\eqref{EqSigmaAqui}. Namely, the key point of the convergence is to be able to find a good approximation of $\bm{\theta}_\texttt{MAP}$ (placing us close to the fixed point). This is possible since we are working in a reduced portion of the complete space, and more efficient Monte Carlo scheme can be applied \cite{StickyLuca,Gilks92,Gilks95,DEEP_IS}. It has the same convergence rate of a  Monte Carlo method for stochastic optimization, as a standard simulated annealing \cite{kirkpatrick83}. 
 {
 	The rate of convergence of an optimizer driven by sampling has been also recently studied in \cite{akyildiz2020parallel, akyildiz2017adaptive}. The authors in \cite{akyildiz2020parallel} show that, in a specific sampler (that iteratively optimizes a posterior pdf), the error bound is  $\mathcal{O}\left(N^{-\frac{1}{2(M+1)}}\right)$, where $M$ is the dimension of the $\btheta$-space.
 }
 \newline
\newline
{\bf Accelerating the convergence of the global optimization problem.}  In other to find a good region of the space for starting the alternating optimization, we can use some iterations (let's say $T_0<T$) of the algorithm considering
	\begin{align}\label{EqMAPAqui_Accelerating}
	\bm{\theta}_\texttt{MAP}^{(t)}  =\arg\min_{\bm{\theta}}  \left[ \sum_{r=1}^R \left|\left|\mathbf{y}_r - {\bf f}_r(\boldsymbol{\theta})\right|\right|^2 - \log g_{\theta}({\bm \theta})\right], \qquad t=1,...,T_0,
\end{align}
that is equivalent to set $\widehat{{\bm \Sigma}}_{\texttt{ML}}^{(t)}={\bf I}_K$ for $ t=0,...,T_0-1$ in Eq. \eqref{EqMAPAqui}, where ${\bf I}_K$ is a $K\times K$ unit matrix.  Thus, in the first $T_0$ iterations, we focus only on finding a good point $\bm{\theta}_\texttt{MAP}^{(T_0)}$. Indeed, note that if there exists a point $\boldsymbol{\theta}^*$ such that  $\sum_{r=1}^R \left|\left|\mathbf{y}_r - {\bf f}_r(\boldsymbol{\theta}^*)\right|\right|^2=0$, then this point  $\boldsymbol{\theta}^*$  is also {\it a root} for $\sum_{r=1}^R\left(\mathbf{y}_r - {\bf f}_r(\boldsymbol{\theta}^*)\right)^{\top}\widehat{{\bm \Sigma}}^{-1} \left(\mathbf{y}_r - {\bf f}_r(\boldsymbol{\theta}^*)\right)=0$ for any possible covariance matrix $\widehat{{\bm \Sigma}}$.
\newline
\newline
{\bf Outputs of this first part of the inference scheme.} With the procedure above, we perform a Bayesian inference  over the vector $\bm{\theta}$, but {\it only} analyzing and approximating the conditional posterior  $p(\bm{\theta}|\obs,\widehat{{\bm \Sigma}}_\texttt{ML}^{(T)})$. In this first part, with respect to ${\bm \Sigma}$, we only provide a frequentist estimator $\widehat{{\bm \Sigma}}_\texttt{ML}^{(T)}$.
\newline
\newline
Note that, in the iterative procedure, we have a sequence conditional posteriors $p(\bm{\theta}|\obs,\widehat{{\bm \Sigma}}_\texttt{ML}^{(t)})$. For this reason, we call the algorithm as  {\it adaptive target adaptive importance sampling} (ATAIS).\footnote{Another reason is that it is also an extension of the techniques in \cite{ATAIS,ATAIS2}, that use the acronym ATAIS as well.}   
The details of the ATAIS algorithm which performs this scheme are given in the next section.

\subsection{Adaptive Target Adaptive Importance Sampling  (ATAIS)}\label{ATAIS_sect}

This section provides more details about Step 1 of the alternating procedure described above.
 More generally, we will provide all the details of the ATAIS algorithm. To simplify the notation, we denote  the unnormalized conditional posterior at the $t$-th iteration, 
\begin{eqnarray}
\pi_t({\bm \theta})=  \ell({\bf Y}|\widehat{{\bm \Sigma}}_\texttt{ML}^{(t-1)},{\bm \theta}) g_\theta({\bm \theta}) \propto p(\bm{\theta}|\obs,\widehat{{\bm \Sigma}}_\texttt{ML}^{(t-1)}).
\label{eq:bayes_3}
\end{eqnarray}
At each iteration, we consider $\pi_t(\bm{\theta})$ as the target distribution. Finally, we are able to approximate $\pi_{T+1}({\bm \theta})  \propto p(\bm{\theta}|\obs,\widehat{{\bm \Sigma}}_\texttt{ML}^{(T)})$, without any additional evaluation of the likelihood function.
The dependence on the iteration $t$ is due to $\widehat{{\bm \Sigma}}_{\texttt{ML}}^{(t)}$ varies with $t$. The ATAIS algorithm is outlined in Table \ref{AIS_AutoTemp}, whereas the main features of ATAIS are described below.
\newline
\newline
{\bf IS steps.}
 A set of $N$ samples $\{{\bm \theta}_{t}^{(n)}\}_{n=1}^N$ are drawn from a (normalized) proposal density $q({\bm \theta}|{\bm \mu}_t,{\bm \Lambda}_t)$ with mean ${\bm \mu}_t$ and a covariance matrix ${\bm \Lambda}_t$. An importance weight 
 $$
 w_{t}^{(n)}=\frac{\pi_t({\bm \theta}_t^{(n)})}{q({\bm \theta}_t^{(n)}|{\bm \mu}_t,{\bm \Lambda}_t)}, 
 $$ 
 is assigned to each sample ${\bm \theta}_t^{(n)}$, for all $n$ and $t$. { Note that we assume that $q({\bm \theta}|{\bm \mu}_t,{\bm \Lambda}_t)$ normalized, i.e., $\int_{\Theta} q({\bm \theta}|{\bm \mu}_t,{\bm \Lambda}_t) d{\bm \theta}=1$ and with heavier tails than $\pi_t$. }
 \newline
\newline
{\bf Proposal adaptation.}  The location parameter of the proposal density is moved to $\widehat{{\bm \theta}}^{(t)}_{\texttt{MAP}}$, i.e.,
\begin{eqnarray}\label{muEqinText}
{\bm \mu}_{t+1}=\widehat{{\bm \theta}}^{(t)}_{\texttt{MAP}}.
\end{eqnarray}
Note that, we set ${\bm \mu}_{t+1}=\widehat{{\bm \theta}}_{\texttt{MAP}}^{(t)}$ instead of using the empirical mean of the samples (as in other classical AIS schemes). This is because we have noticed that this choice provides better and more robust results, especially as the dimension of the problem grows.  Indeed, this choice helps in the search for the global maximum (since the next cloud of particles will be around the current MAP estimation) and, as a consequence, helps also the estimation of $\widehat{{\bm \Sigma}}_\texttt{ML}$ due to \eqref{EqSigmaAqui}. {   Regarding the adaptation of  the covariance matrix ${\bm \Lambda}_t$ of the proposal density, see Section \ref{SectComp}.}
 \newline
 \newline
{\bf ATAIS outputs.} 
After $T$ iterations, a final correction of the weights is needed, i.e.,
\begin{equation}\label{aquiW}
\widetilde{w}_{t}^{(n)}=w_{t}^{(n)} \frac{\pi_{T+1}({\bm \theta}_t^{(n)})}{\pi_t({\bm \theta}_t^{(n)})}, \qquad \mbox{for all $n,t$},
\end{equation}
in order to obtain a particle approximation of the measure of the final conditional posterior $\pi_{T+1}(\bm{\theta})\propto p(\bm{\theta}|\obs,\widehat{{\bm \Sigma}}_\texttt{ML}^{(T)})$. Thus, the algorithm  returns the final estimators $\widehat{{\bm \theta}}_{\texttt{MAP}}^{(T)}$, $\widehat{{\bm \Sigma}}_{\texttt{ML}}^{(T)}$, and all the weighted samples $\{{\bm \theta}_{t}^{(n)},\widetilde{w}_{t}^{(n)}\}$, for all $n=1,...,N$ and $t=1,...,T$. Other outputs can be obtained with a post-processing of the weighted samples, as shown below. Note that Eq. \eqref{aquiW} does not require any additional evaluations of the model, if we save the computation of the error vectors  ${\bf e}_{t,r}^{(n)}={\bf y}_r- {\bf f}_r({\bm \theta}_t^{(n)})$.
Moreover, we can also use $\{{\bf e}_{t,r}^{(n)}\}$  and $\{{\bm \theta}_{t}^{(n)}\}$ for building a particle approximation of  any other conditional posterior $p(\bm{\theta}|\obs, {\bm \Sigma})${, as proposed in the second part of the inference. 
Note that ATAIS is an adaptive importance sampler. Alternatively, if we would like to employ  an adaptive MCMC algorithm in Table \ref{AIS_AutoTemp} (instead of an importance sampler), - assuming we use a sufficiently large $T$ - we should set $w_{t}^{(n)}=1$ in Eq. \eqref{aquiW}. In any case, the weights would be $\widetilde{w}_{t}^{(n)}= \frac{\pi_{T+1}({\bm \theta}_t^{(n)})}{\pi_t({\bm \theta}_t^{(n)})}$ in Eq. \eqref{aquiW}, keeping still the IS nature. }


\subsection{Elliptically contoured distributions in the observation models} \label{OtherNoiseSect}
The ATAIS algorithm described above (including the alternating optimization above) can be also applied for more general models. When the noise has an elliptically contoured distribution \cite{Dagpunar88,MartinoBook,Fang01}, i.e.,
\begin{align}\label{NoiseVariable}
\mathbf{v}_r\sim p(\mathbf{v}), \quad \mbox{ with } \quad p(\mathbf{v})=\frac{k_p}{|\boldsymbol{\Sigma}|^{1/2}} h\left(\mathbf{v}^{\top} \boldsymbol{\Sigma}^{-1}\mathbf{v}\right),
\end{align}
where $k_p>0$ is a constant, $h(s): \mathbb{R} \rightarrow \mathbb{R}^{+}$is a one-dimensional positive function. The $K\times K$ matrix ${\bm \Sigma}$ is  a scale matrix related to the corresponding covariance matrix. An example is the  multivariate $t$-distribution:
$$
p(\mathbf{v})\propto\frac{1}{|\boldsymbol{\Sigma}|^{1/2}} \left[1+\frac{1}{\nu}\mathbf{v}^\top {\bm \Sigma}^{-1}\mathbf{v}\right]^{-(\nu+K) / 2},
$$
where $\nu>0$ represents the degrees of freedom.
{
However, {unlike in the Gaussian case, in this scenario the empirical covariance matrix is not an  estimator of the scale matrix $\boldsymbol{\Sigma}$, but of a scaled version of the scale matrix, i.e., $\alpha\bSigma$ where $\alpha>0$.  } Generally, the estimator depends on the specific function $h(s)$  \cite{bilodeau1999theory}.\footnote{In \cite[Chapter 13]{bilodeau1999theory}, the interested reader can find the properties that the elliptical distribution must satisfy for the existence and uniqueness of the estimators of the location and scale parameters.} 
 Let us consider the problem of estimating only the scale parameter keeping fixed ${\bf f}({\bm \theta})$, as in step \ref{EsteSTEP_Sigma} of ATAIS. When the conditions expressed in \cite{bilodeau1999theory} are satisfied (which are satisfied for absolutely continuous distributions such as Student's-t densities), the scaling matrix $\bSigma$ can be estimated by a fixed-point method as follows.
Let define an initialization $\widehat{\bSigma}_0>0$, the function $\eta(s) = -2 \dfrac{h'(s)}{h(s)}$ and ${\bf e}_r = \textbf{y}_r - {\bf f}_r(\btheta)$, with $r=1,\dots, R$. We estimate $\bSigma$ by the iterative formula:
\[ 
\widehat{{\bm \Sigma}}_k = \dfrac{1}{R} \sum_{r=1}^{R} \eta\left({\bf e}_r^{\top} \left[\widehat{{\bm \Sigma}}_{k-1}\right]^{-1} {\bf e}_r\right) {\bf e}_r{\bf e}_r^{\top}, 
\]
where $k$ represents the index of this iterative process.
Note that $\eta(s)$ depends on $h(s)$. More sophisticated procedures for estimating $\bSigma$  can be found in \cite{Hediger_2023, ke2019user}. Finally, we remark that  an ATAIS scheme could be also designed when the noise has a  generalized Gaussian density. In one dimension, this pdf has the form $p(v)=\frac{\beta}{2 s \Gamma(1 / \beta)} \exp\left(-\frac{|v-\mu|^\beta}{s^\beta}\right)$  where, fixing $\mu$, $\beta$, we know the estimator ${\widehat s}=\left(\frac{\beta}{N} \sum_{i=1}^N\left|x_i-\mu\right|^\beta\right)^{1 / \beta}$.



{
\section{Speeding up, robustness and possible alternatives}\label{SectComp}

\subsection{About the choice of the type of proposal density, $N$ and $T$}

Regarding the choice of the proposal, we need to be able to evaluate it pointwise (it must be normalized) and be able to draw samples from it.  
The other additional requirement is that the proposal density must have heavier tails than the target pdf, i.e., the conditional posterior $\pi_t({\bm \theta})$. 
The presence of multimodality in the target pdf is not an issue, especially if the cyclic adaptation of the covariance matrix of the proposal is employed (see Section \ref{AdapCOVprop}). Moreover, multiple proposal schemes can be used, to deal with multimodality (see Section \ref{AdMISaqui}).
\newline
The consistency of the ATAIS estimators is ensured when $N\rightarrow \infty$ (with finite $T$) or $T\rightarrow \infty$ (with finite $N$).  However, at least theoretically, in any adaptive importance sampler the choice of the value of $N$ is more critical than $T$, since the adaptation of the proposal pdf is driven by the $N$ samples for each iteration. In ATAIS, the value $N$ is even more important, since the conditional posterior is also adapted (due to the estimation and update of  $\widehat{{\bm \Sigma}}_\texttt{ML}^{(t)}$). Namely, the $N$ weighted samples per iteration affect (a) the proposal adaptation and also (b) the first step of the alternating optimization in Table \ref{TableAO} i.e., more specifically, in step \ref{StepMaxIter} of Table \ref{AIS_AutoTemp}. However, the simulations in Section \ref{SectMO} and Figure \ref{Tu_puta_madre}  show the importance of the adaptation in ATAIS and as a consequence of the number of iterations $T$. 

\subsection{Adaptation of the covariance matrix of the proposal density}\label{AdapCOVprop}
The adaptation of the parameters of the proposal density is described in Eqs. \eqref{muEqinText}-\eqref{muEqTable} and  \eqref{SigmaEqTable} of Table \ref{AIS_AutoTemp}. Regarding the covariance matrix ${\bm \Lambda}_t$  we consider Eq. \eqref{SigmaEqTable} that is formed by two parts as follows:  
\begin{eqnarray}\label{SigmaEqNoTable}
{\bm \Lambda}_{t+1}&=&\underbrace{\sum_{n=1}^N \bar{w}_t^{(n)} ({\bm \theta}_t^{(n)}-{\bar {\bm \theta}}_t)^{\top} ({\bm \theta}_t^{(n)}-{\bar {\bm \theta}}_t)}_{\mbox{empirical}} + \underbrace{\delta_t {\bf I}_M}_{\mbox{cyclic}}, \qquad \delta_t >0.
\end{eqnarray}
The first term is the empirical covariance of the weighted samples at the $t$-th iteration. However, this term can be harmful to the performance, especially at the beginning of the adaptation, often yielding to empirical covariance matrices with very small traces (depending on the degeneracy of the weighted samples). Robust alternatives to this first term are discussed in \cite{ElLaham2018RobustCA}.
\newline
The second term is a diagonal matrix controlled by a positive parameter $\delta_t>0$ which determines the elements in the diagonal. This term helps the IS performance (see , e.g., \cite[Numerical Example 1]{Llorente20}) and avoids catastrophic scenarios. Moreover, this second term always ensures to have positive definite matrices ${\bm \Lambda}_t$ and avoid numerical issues.
 Generally,  one should use a greater value of $\delta_t$ especially in the first iterations, and the value of $\delta_t$ could be decreased as the iterations grow. Generally, one should use a large value of $\delta_t$ in the initial iterations, and gradually decrease the value of $\delta_t$ as the iterations progress.  For a robust implementation, here we suggest to apply a cyclic/periodic strategy, i.e., 
 \begin{gather}\label{DeltStrategy}
\delta_{t+1}=  \left\{
  \begin{split}
&a \ \delta_{t}, \qquad   \mbox{ if }\delta_t \geq  \delta_{\texttt{min}}, \\
& \delta_{0},    \mbox{ }\mbox{ }\mbox{ }\qquad \mbox{ if }\delta_t < \delta_{\texttt{min}},
 \end{split}
 \right.
  \end{gather}
  with  $0< a <1$, where $\delta_0$ is the maximum value that the user desires to use for exploration (that is also an initial value) and $\delta_{\texttt{min}}>0$ is the minimum possible value that can be employed avoiding numerical problems.
 The goal is to combine iterations devoted to the {\it exploration} (associated with greater values of $\delta_t$,  close to $\delta_0$) and other iterations devoted to the {\it exploitation}  (associated with smaller values of $\delta_t$,  close to $\delta_{\texttt{min}}$). The coefficient $a$ determines the speed of decrease of $\delta_t$ and, as a consequence, the number of cycles/periods. 
Note that, in any case, the first term which depends on the weighted samples can drive the final covariance matrix ${\bm \Lambda}_t$.

\subsection{Saving and re-weighting only few particles}

The first part of ATAIS yields $NT$ weighted samples. These $NT$ samples must be re-weighted in the last step as  in Eq. \eqref{aquiW} or \eqref{EQ_AQUIreweighting}. 
At each iteration, we should save $N$ weighted particles. This can produce an important computational load. However, most of these samples have a negligible weight, especially in the first iterations, when ATAIS is still looking for the regions of high probability.  The use of the resampling procedure (i.e., bootstrap) over $N$ samples at each iteration or $NT$ final samples is not advisable, since (if $N$ large) it can be very slow.
\newline
We suggest to save, at each iteration $t$, only the (extremely) relevant samples with a normalized weight $\bar{w}_t^{(n)}\geq\frac{1}{N}$. This criterion is also connected with a possible effective sample size (ESS) measure proposed and discussed in \cite{ESSarxiv16}. Note that if only a few samples have a non-negligible weight (i.e., the - very frequent - degenerate case), we just save the relevant ones. On the other hand, if all samples have similar weights (and, as a consequence, all the normalized weights $\bar{w}_t^{(n)}$ are around $1/N$), this is owing to two possible scenarios:
\begin{itemize}
\item All the samples are located in a tail of distribution (an almost-flat region that yields similar weights). In this case, most of the samples are not relevant, and we do not  lose information if saving only some of them.  
\item The samples are all relevant, and located in high probability regions. In this case, we lose some information (saving only some of them) but, since the proposal density is already well-adapted (i.e., well-located), in the next iterations we will draw more and more relevant samples around these high probability regions. 
\end{itemize}
Moreover, for the second part of the inference in Section \ref{SectSuperFer}, we need to save only the computation of the error vectors ${\bf e}_{t,r}^{(n)}={\bf y}_r- {\bf f}_r({\bm \theta}_t^{(n)})$, e.g., see Eq. \eqref{eq:LH_2}. Thus, 
no additional evaluation of the model are required in this second part. Hence, at each iteration, we suggest just  to save  the error vectors ${\bf e}_{t,r}^{(n)}$ of the particles with a normalized weight such that $\bar{w}_t^{(n)}\geq\frac{1}{N}$.


\subsection{Optimal denominator in the IS weights}
 Since we adapt the proposal density during the iterations, we are actually in a multiple IS scenario \cite{ElviraMIS15,Llorente22mcmcDriven}.  It is well-known that the standard IS denominator (using just the unique proposal $q(\theta|{\bm \mu}_t, {\bm \Lambda}_t)$) provides instability and high variance in the final IS estimators. The correct way of avoiding this behavior is to employ a mixture of all proposals used during the iterations, i.e.,
\begin{align}\label{MISweights}
w_t^{(n)} = \frac{\pi_t(\bm{\theta}_t^{(n)})}{\frac{1}{t}\sum_{i=1}^{t} q(\bm{\theta}_t^{(n)}| {\bm \mu}_i, {\bm \Lambda}_i )}.
\end{align}
This procedure provides the lowest variance of the final IS estimators but requires a high computational cost. Indeed, for each sample $\bm{\theta}_t^{(n)}$, we have to evaluate a mixture where the number of components grows with the iterations. Moreover, {\it at least} in the final iteration $T$  decided by the user, all the  previous weights must be updated recomputing a new denominator for each sample. Alternatives for reducing the computational cost have been proposed \cite{EUSIPCO19EffAMIS}. The simplest solution among the proposed ones is to build a {\it compressed} denominator \cite{CMCpaper,Llorente22mcmcDriven}. To avoid instabilities in the results, we can discard the samples and the proposals in the first iterations (hence, we do not use the means ${\bm \mu}_t=\widehat{{\bm \theta}}^{(t)}_{\texttt{MAP}}$), when the parameters of the proposal density change substantially. For instance, one can discard the corresponding means ${\bm \mu}_t=\widehat{{\bm \theta}}^{(t)}_{\texttt{MAP}}$ in the first iterations $t$ such that $\frac{||\widehat{{\bm \theta}}^{(t)}_{\texttt{MAP}}-\widehat{{\bm \theta}}^{(t-1)}_{\texttt{MAP}} ||}{\max\left[||\widehat{{\bm \theta}}^{(t)}_{\texttt{MAP}}||,||\widehat{{\bm \theta}}^{(t-1)}_{\texttt{MAP}}||\right]}>\epsilon$ where $\epsilon�\in (0,1]$. By our empirical experience, we suggest the use of $\epsilon=0.3$. However, we remark that the use of the weights of type \eqref{MISweights} is not mandatory. ATAIS provides very good performance also with the standard IS  weights, as shown by the numerical experiments in Section \ref{sec:Simul}. 

\subsection{Possible use of multiple proposal densities}\label{AdMISaqui}

In order to foster the robustness of the algorithm, one possibility is to use more than one proposal density at each iteration. The simplest way for extending ATAIS with $H$ multiple proposals is described next. Denoting each proposal as $q_h({\bm \theta}|{\bm \mu}_{t,h},{\bm \Lambda}_{t,h})$ with $h=1,...,H$, the idea is to draw $N$ samples for each proposal pdf,
$$
{\bm \theta}_{t,h}^{(1)},...,{\bm \theta}_{t,h}^{(N)} \sim q_h({\bm \theta}|{\bm \mu}_{t,h},{\bm \Lambda}_{t,h}), \qquad h=1,...,H,
$$
and weigh them keeping a {\it unique} (adaptive) posterior distribution $\pi_t({\bm \theta})= \ell({\bf Y}|\widehat{{\bm \Sigma}}_\texttt{ML}^{(t-1)},{\bm \theta}) g_\theta({\bm \theta})$, as
\begin{eqnarray}
w_{t,h}^{(n)}=\frac{\pi_t({\bm \theta}_{t,h}^{(n)})}{q_h({\bm \theta}_{t,h}^{(n)}|{\bm \mu}_{t,h},{\bm \Lambda}_{t,h})}, \qquad \mbox{ or } \qquad w_{t,h}^{(n)}=\frac{\pi_t({\bm \theta}_{t,h}^{(n)})}{\frac{1}{H} \sum_{j=1}^H q_j({\bm \theta}_{t,j}^{(n)}|{\bm \mu}_{t,j},{\bm \Lambda}_{t,j})},
\end{eqnarray}
where, in the second option, we have employed the multiple IS approach \cite{ElviraMIS15,Llorente22mcmcDriven}, similarly as in Eq. \eqref{MISweights}.
 Then, the idea is to perform an independent  adaptation of the mean (as in \cite{APIS15}) considering the local MAP estimations, i.e.,
\begin{eqnarray}
{\bm \mu}_{t+1,h}=\widehat{{\bm \theta}}^{(t,h)}_{\texttt{MAP}}, 
\end{eqnarray}
where we consider only the samples generated by the $m$-th proposal. The same approach can be employed for adapting ${\bm \Lambda}_{t,m}$, i.e., independently to each other (see Section \ref{AdapCOVprop}). Hence, in this sense, the steps described so far could be parallelized.
Now, denoting as 
\begin{align}
\overline{{\bm \theta}}^{(t)}_{\texttt{MAP}}=\arg\max\limits_{h} \pi_t(\widehat{{\bm \theta}}^{(t,h)}_{\texttt{MAP}}),
\end{align}
the best MAP estimator at the t-$th$ iteration, we can obtain a {\it unique} estimation of the covariance matrix, 
	\begin{align}\label{EqSigmaAqui_otravez}
\widehat{{\bm \Sigma}}_{\texttt{ML}}^{(t)}=\frac{1}{R} \sum_{r=1}^R\left({\bf y}_r- {\bf f}_r({\overline{\bm \theta}}^{(t)}_{\texttt{MAP}})\right)\left({\bf y}_r- {\bf f}_r({\bm {\overline\theta}}^{(t)}_{\texttt{MAP}})\right)^{\top},
\end{align}
so that we can have a unique posterior  $\pi_{t+1}({\bm \theta}) =  \ell({\bf Y}|\widehat{{\bm \Sigma}}_\texttt{ML}^{(t)},{\bm \theta}) g_\theta({\bm \theta})$ for the next iteration of the algorithm. In summary, all the proposal pdfs are adapted independently but  they face the same posterior. The rest of the algorithm remains unaltered as in Table \ref{AIS_AutoTemp}. More sophisticated adaptive strategies could include a ``repulsion'' among the proposals, as suggested in \cite{ELVIRAGrad,MarParChain0}.


 



\subsection{Possible use of mini-batches}\label{BatchSect} 
When the number of vectors ${\bf y}_r$ of data $R$ grows, different issues appear: 
\begin{itemize}
\item  the calculation of the likelihood becomes costly/slow;
 \item numerical problems due to the finite precision of the employed machine;
 \item the exploration of the high probability regions becomes more difficult since the posterior is extremely tight. 
\end{itemize}
All these issues are mitigated by using mini-batches of data. 
 ATAIS allows the direct use of mini-batches (see \cite{AdaptiveNoisyImportanceSamplingForStochasticOptimization,NoisyISFernando}). Namely, we can use a sub-set of data (e.g., formed by $L<R$ vectors ${\bf y}_i$ of data) to create sub-posteriors, 
\begin{equation}\label{equ:subposterior}
\widetilde{\pi}_t({\bm \theta})\propto \left[\prod_{i=1}^{L} \ell({\bf y}_{k_i} | {\bm \Sigma}_\texttt{ML}^{(t-1)}, \bm{\theta}) \right] g_{\bm{\theta}}(\bm{\theta}),  
\end{equation}
where  ${\bf y}_{k_i} \in \{{\bf y}_{1},...,{\bf y}_{R}\}$, i.e., $k_i\in\{1,...,R\}$ with $i=1,...,L$, are $L<R$ vectors that can be selected randomly over the $R$ possible vectors \cite{AdaptiveNoisyImportanceSamplingForStochasticOptimization,NoisyISFernando}. 
ATAIS can use the subposteriors, defined as in Eq. \eqref{equ:subposterior}, in each step \ref{StepMaxIter} of Table \ref{AIS_AutoTemp}, i.e.,
\begin{align}
{\bm \theta}^{(t)}_{\texttt{max}} =\arg\max\limits_{n} \widetilde{\pi}_t({\bm \theta}_t^{(n)}), \qquad \widehat{\bf{r}}_t={\bf f}_r({\bm \theta}^{(t)}_{\texttt{max}}), \qquad \widehat{{\bm \Sigma}}_t=\frac{1}{L} \sum_{j=1}^{L} (\bm{y}_j - \widehat{\bf{r}}_t) (\bm{y}_j - \widehat{\bf{r}}_t)^{\top},
\end{align}
 using only the $L$ selected data vectors. Below, we describe two possible schemes for modifying ATAIS with mini-batches.
 \newline
 \newline
{\bf First possible strategy.} In the standard ATAIS, we can perform the check in step \ref{GlobCheck}, obtaining a sequence of 
$\widehat{{\bm \theta}}_{\texttt{MAP}}^{(t)}$ and $ \widehat{{\bm \Sigma}}_\texttt{ML}^{(t)}$.
Clearly, the proposal density $q({\bm \theta}|{\bm \theta}_{\texttt{MAP}}^{(t)},{\bm \Lambda}_t)$ is adapted as in Eqs. \eqref{muEqTable}-\eqref{SigmaEqTable}. 
After all the $T$ iterations, we consider the computation of the complete conditional posterior in Eq. \eqref{eq:bayes_3}  (with {\it all} the data) is feasible (at least for a few points). We can compute 
$$
\pi_t(\widehat{{\bm \theta}}_{\texttt{MAP}}^{(t)}) \propto p({\bm \theta}_{\texttt{MAP}}^{(t)}|\widehat{{\bm \Sigma}}_\texttt{ML}^{(t)},{\bf Y}),
$$ 
and 
\begin{eqnarray}
 \overline{{\bm \theta}}_{\texttt{MAP}}=\arg\max \left\{ \pi_t\left(\widehat{{\bm \theta}}_{\texttt{MAP}}^{(t)}\right), \mbox{ for all $t=1,...,T$}\right\}.
 \end{eqnarray}
The  final estimation of the matrix can be done 
\begin{eqnarray}
\overline{{\bm \Sigma}}_{\texttt{ML}}=\frac{1}{R} \sum_{r=1}^R\left({\bf y}_r- {\bf f}_r({\overline{\bm \theta}}_{\texttt{MAP}})\right)\left({\bf y}_r- {\bf f}_r({\bm {\overline\theta}}_{\texttt{MAP}})\right)^{\top},
 \end{eqnarray}
considering all the $R$ data vectors. 
Moreover, in the second part of the proposed inference scheme (see below in the next section),   the final re-weighting step must be made according to the full posterior, so that the final estimations are performed considering all the dataset. Hence, in this strategy, the complete ATAIS method can consider the use of mini-batches only in the $T$ iterations of the first part, whereas, after the $T$ iterations for computing above and in the second part, the full-posterior must be evaluated.
\newline
{\bf Second possible strategy.} Otherwise, if the computation of the complete conditional posteriors is too costly even for a few points, we can follow the approach suggested in \cite{embarrassinglyparallel}. 
 Let us consider the following assumptions:
 \begin{itemize}
\item  Assume that all the data are divided into $\frac{R}{L}$ disjoint subsets (so that the union of these subsets includes all the data) so that the complete posterior can be expressed as the product of the $\frac{R}{L}$ sub-posteriors, and to select the total number of iterations as  $T=\frac{R}{L}$.
\item Assume that, at least after a certain iteration, the variations in $\widehat{{\bm \Sigma}}_{\texttt{ML}}^{(t)}$ are small,  so that $\widehat{\pi}_{t-1}({\bm \theta})  \approx \widehat{\pi}_t({\bm \theta}) \approx \widehat{\pi}_{t+1}({\bm \theta})...$ and so on, i.e., the conditional full-posterior is virtually not changing. This assumption is verified when $\widehat{{\bm \theta}}_{\texttt{MAP}}^{(t)}$  is close to  the true vector ${\bm \theta}_{\texttt{MAP}}$.
\item  Assume as prior in each sub-posterior, $g_{\texttt{sub}}({\bm \theta})=g({\bm \theta})^{1/T}$.
\end{itemize}
Then, the conditional full-posterior (that virtually is not changing) can be expressed as the product of the $T=\frac{R}{L}$ (conditional) sub-posteriors.
 At each iteration of ATAIS,
we can consider a Gaussian approximation of each sub-posterior, $\widehat{p}_t({\bm \theta})=\mathcal{N}({\bm \theta}|{\bm \theta}^{(t)}_{\texttt{max}},{\bm \Lambda}_t)$ (with $\delta_t$ in Section \ref{AdapCOVprop} is small, close to zero), with mean ${\bm \theta}^{(t)}_{\texttt{max}}$  and covariance matrix ${\bm \Lambda}_t$, both computed according to the sub-posterior $\widetilde{\pi}_t({\bm \theta})$ in Eq. \eqref{equ:subposterior} considering the current mini-batch.
  Hence an approximation of conditional full-posterior can be also a Gaussian distribution,
 \begin{align}
\widehat{\pi}_t({\bm \theta})=\prod_{\tau}^t \widehat{p}_t({\bm \theta}) \propto \mathcal{N}\left({\bm \theta} \mid  \widehat{{\bm \theta}}_{\texttt{MAP}}^{(t)},{\bm \Lambda}_{\texttt{tot}}^{(t)}\right),
\end{align}
 with covariance matrix
 \begin{align}
{\bm \Lambda}_{\texttt{tot}}^{(t)}=\left(\sum_{\tau=1}^{t} {\bm \Lambda}_{\tau}^{-1}\right)^{-1}, 
\end{align}
and mean
\begin{align}
 \widehat{{\bm \theta}}_{\texttt{MAP}}^{(t)}=  {\bm \Lambda}_{\texttt{tot}}\sum_{\tau=1}^{t} {\bm \Lambda}_{\tau}^{-1}{\bm \theta}^{(\tau)}_{\texttt{max}}.
\end{align}
Note that, we can also have
\begin{eqnarray}
\widehat{{\bm \Sigma}}_{\texttt{ML}}^{(t)}=\frac{1}{R} \sum_{r=1}^{R}\left({\bf y}_r- {\bf f}_r( \widehat{{\bm \theta}}_{\texttt{MAP}}^{(t)})\right)\left({\bf y}_r- {\bf f}_r( \widehat{{\bm \theta}}_{\texttt{MAP}}^{(t)})\right)^{\top}.
 \end{eqnarray}
Therefore, we can replace the check in step \ref{GlobCheck} of Table \ref{AIS_AutoTemp} with the computation of $\widehat{{\bm \theta}}_{\texttt{MAP}}^{(t)}$ above $\widehat{{\bm \Sigma}}_{\texttt{ML}}^{(t)}$. 
Note that, unlike the proposal $q({\bm \theta}|{\bm \theta}_{\texttt{MAP}}^{(t)},{\bm \Lambda}_t)$ is still adapted as in Eqs. \eqref{muEqTable}-\eqref{SigmaEqTable}. 
The final Gaussian approximation with $\widehat{{\bm \theta}}_{\texttt{MAP}}^{(T)}$, ${\bm \Lambda}_{\texttt{tot}}^{(T)}$ and the final estimation $\widehat{{\bm \Sigma}}_{\texttt{ML}}^{(T)}$, can be also employed in the second part of the inference scheme instead of evaluating the true full-posterior. Thus, in this second approach,  we never need the evaluation of the (true) full-posterior. 

\subsection{Possible use of the gradient information if the likelihood is differentiable} 
 The current version of ATAIS can be applied even when the model ${\bf f}({\bm \theta})$ is not differentiable. Moreover, the Monte Carlo approach ensures the convergence to the global maximum as $N\rightarrow \infty$ or $T \rightarrow \infty$. However, if ${\bf f}({\bm \theta})$ is differentiable and the practitioner desires to incorporate the gradient information several alternatives strategies can be employed.  The simplest idea is to replace the step \ref{StepMaxIter} of  Table  \ref{AIS_AutoTemp}, by a gradient descent step over the function 
 $$
 C({\bm \theta})=-\log \pi_t({\bm \theta})=  -\log \ell({\bf Y}|\widehat{{\bm \Sigma}}_\texttt{ML}^{(t-1)},{\bm \theta}) g_\theta({\bm \theta}),
 $$
  i.e.,
$$
{\bm \theta}^{(t)}_{\texttt{max}}={\bm \theta}^{(t-1)}_{\texttt{MAP}}-\gamma \nabla C\left({\bm \theta}^{(t-1)}_{\texttt{MAP}}\right) +e_t,
$$
with $\gamma>0$ and $e_t$ is a noise perturbation, avoiding to remain stuck in a local minimum. Thus,  ${\bm \theta}^{(t+1)}_{\texttt{max}}$ would be a candidate as a possible new mean of the proposal pdf at step \ref{GlobCheck} of Table \ref{AIS_AutoTemp}, i.e., ${\bm \mu}_t=\widehat{{\bm \theta}}^{(t)}_{\texttt{MAP}}={\bm \theta}^{(t)}_{\texttt{max}}$, if $\pi_t\big({\bm \theta}^{(t)}_{\texttt{max}}\big) > \pi_{\texttt{MAP}}$, (where $\pi_{\texttt{MAP}}$ is the greatest value so far obtained in Eq. \eqref{eq:bayes_3}, considering also the corresponding estimated covariance matrix $\widehat{{\bm \Sigma}}$ is also changing). See \cite{rodgers2000inverse}, for all the formulas required in the case of Gaussian noise perturbation, as in Eq. \eqref{EqGaussNoise}.
Clearly, it  is also possible to use a stochastic gradient descent using mini-batches \cite{StoGradMinBatches}.
More sophisticated gradient-based IS schemes  have been discussed in literature \cite{ELVIRAGrad}, also involving multiple proposal densities and repulsion \cite{ELVIRAGrad,schuster2015gradientimportancesampling,FasioloGrad, rodgers2000inverse}. 
}
%

}

\section{Second part of the proposed inference scheme}\label{SectSuperFer}

Here, we describe the second part of the ATAIS procedure, which allows a complete Bayesian analysis of ${\bm \theta}$ and ${\bm \Sigma}$.  It is important to remark that this second part of ATAIS does not require any additional sample generation and likelihood evaluation. Indeed, ATAIS recycles and reweights the samples ${\bm \theta}_t^{(n)}$  obtained in the first part by Table \ref{AIS_AutoTemp}.

\subsection{Approximating different conditional posteriors}\label{DiffCondPost}
 The idea here is to re-use all the generated samples since if we have saved the computation of the error vectors ${\bf e}_{t,r}^{(n)}={\bf y}_r- {\bf f}_r({\bm \theta}_t^{(n)})$
no additional evaluation of the model are required. Note that  the cloud of particles $\{{\bm \theta}_t^{(n)}\}$ is well-located since ATAIS works to generate samples around the MAP and ML estimators of ${\bm \theta}$ and ${\bm \Sigma}$. Moreover, we can also use $\{{\bf e}_{t,r}^{(n)}\}$  and $\{{\bm \theta}_{t}^{(n)}\}$ for building a particle approximation of  any other conditional posterior $p(\bm{\theta}|\obs, {\bm \Sigma})$, i.e., 
\begin{align}\label{FinalParApprox2}
\widehat{p}(\bm{\theta}|\obs,{\bm \Sigma})=\sum_{t=1}^T \sum_{n=1}^N \bar{\rho}_{t}^{(n)}({\bm \Sigma})  \cdot \delta({\bm \theta}-{\bm \theta}_t^{(n)}), \qquad\quad \displaystyle\sum_{t=1}^T\sum_{n=1}^N  \bar{\rho}_{t}^{(n)}({\bm \Sigma}) =1,
\end{align}
where
\begin{align}\label{SuperEqAqui1}
 \rho_{t}^{(n)}({\bm \Sigma})&=\frac{\ell({\bf Y}|{\bm \theta}_t^{(n)},{\bm \Sigma})g_\theta({\bm \theta}_t^{(n)})}{q({\bm \theta}_t^{(n)}|{\bm \mu}_t,{\bm \Lambda}_t)}, \quad \mbox{ and }  \\
\bar{\rho}_{t}^{(n)}({\bm \Sigma})&=\frac{\rho_{t}^{(n)}({\bm \Sigma})}{\sum_{\tau=1}^T\sum_{i=1}^N  \rho_{\tau}^{(i)}({\bm \Sigma})}. \label{SuperEqAqui2}
\end{align}
Given a new matrix ${\bm \Sigma}$,  to compute the likelihood 
\begin{eqnarray}
 \ell({\bf Y}|{\bm \theta}_t^{(n)}, {\bm \Sigma})&=&
 \left(\frac{1}{(2\pi)^{K/2} \mbox{det}({\bm \Sigma})^{1/2}}\right)^R \exp\left( -\frac{1}{2}  
\sum_{r=1}^R\left(\mathbf{y}_r - {\bf f}_r\left(\boldsymbol{\theta}_t^{(n)}\right)\right)^{\top}{\bm \Sigma}^{-1} \left(\mathbf{y}_r - {\bf f}_r\left(\boldsymbol{\theta}_t^{(n)}\right)\right)
 \right),  \\
 &=&
 \left(\frac{1}{(2\pi)^{K/2} \mbox{det}({\bm \Sigma})^{1/2}}\right)^R \exp\left( -\frac{1}{2} 
\sum_{r=1}^R \left({\bf e}_{t,r}^{(n)}\right)^{\top}{\bm \Sigma}^{-1} \left({\bf e}_{t,r}^{(n)}\right),
   \right) 
\label{eq:LH_2}
\end{eqnarray}
we need the vectors ${\bf e}_{t,r}^{(n)}$, the inverse matrix of ${\bm \Sigma}$ and the determinant of ${\bm \Sigma}$.

\subsection{Approximation of the complete posterior distribution and marginal likelihood}

We can apply an IS scheme with the complete target pdf,
\begin{align}\label{completePostEQ}
p(\bm{\theta},{\bm \Sigma} |\obs)=\frac{p(\bm{\theta},{\bm \Sigma} ,\obs)}{p(\obs)} =\frac{\ell({\bf Y}|{\bm \theta}, {\bm \Sigma}) g_\theta({\bm \theta}) g_{\bm \Sigma}( {\bm \Sigma})}{p(\obs)}, \\
 \propto  \ell({\bf Y}|{\bm \theta}, {\bm \Sigma}) g_\theta({\bm \theta}) g_{\bm \Sigma}( {\bm \Sigma}),
\end{align}
 and employing a proposal density that can be factorized as $q({\bm \theta}|{\bm \mu}_t,{\bm \Lambda}_t) q_{\bm \Sigma}( {\bm \Sigma})$ where the piece of proposal $q({\bm \theta}|{\bm \mu}_t,{\bm \Lambda}_t)$ is the same used in ATAIS at the $t$-th iteration. Recycling the $NT$ samples produced by ATAIS, i.e., ${\bm \theta}_t^{(n)} \sim q({\bm \theta}|{\bm \mu}_t,{\bm \Lambda}_t)$ and drawing $J$ random matrices from the proposal $q_{\bm \Sigma}( {\bm \Sigma})$, i.e.,  ${\bm \Sigma}^{(j)} \sim q_{\bm \Sigma}( {\bm \Sigma})$, the complete IS weights are
\begin{align}
\beta_{t,j}^{(n)}&=\frac{\ell({\bf Y}|{\bm \theta}_t^{(n)}, {\bm \Sigma}^{(j)}) g_\theta({\bm \theta}_t^{(n)}) g_{\bm \Sigma}( {\bm \Sigma}^{(j)})}{q({\bm \theta}_t^{(n)}|{\bm \mu}_t,{\bm \Lambda}_t) q_{\bm \Sigma}( {\bm \Sigma}^{(j)})}, \\
&=\rho_{t}^{(n)}({\bm \Sigma}^{(j)}) \frac{ g_{\bm \Sigma}( {\bm \Sigma}^{(j)})}{ q_{\bm \Sigma}( {\bm \Sigma}^{(j)})} =\rho_{t}^{(n)}({\bm \Sigma}^{(j)}) \gamma_j= \rho_{t,j}^{(n)}\gamma_j,
\label{AquiGammaW}
\end{align}
where we have set $ \gamma_j=\frac{ g_{\bm \Sigma}( {\bm \Sigma}^{(j)})}{ q_{\bm \Sigma}( {\bm \Sigma}^{(j)})}$ and $\rho_{t,j}^{(n)}=\rho_{t}^{(n)}({\bm \Sigma}^{(j)})$ are given in Eq. \eqref{SuperEqAqui1}. Clearly, if $q_{\bm \Sigma}( {\bm \Sigma})= g_{\bm \Sigma}( {\bm \Sigma})$ then $ \gamma_j=1$. The complete posterior approximation is given by
\begin{align}
 \widehat{p}(\bm{\theta}, {\bm \Sigma}|\obs)=
\sum_{j=1}^J  \sum_{t=1}^T \sum_{n=1}^N \bar{\beta}_{t,j}^{(n)} \cdot \delta({\bm \theta}-{\bm \theta}_t^{(n)}) \delta\left({\bm \Sigma}-{\bm \Sigma}^{(j)}\right),
\end{align}
where $ \bar{\beta}_{t,j}^{(n)}=\frac{\beta_{t,j}^{(n)}}{\sum_{i=1}^J  \sum_{v=1}^T \sum_{m=1}^N\beta_{v,i}^{(m)}}$.
Note that we have different numbers of samples about ${\bm \theta}$ (i.e., $NT$) and ${\bm \Sigma}$ (i.e., $J$). This recalls the recycling Gibbs sampling idea in \cite{RecGibbs}, where the space is divided into blocks, and different numbers of samples are considered for each block.
\newline 
The marginal likelihood $p(\obs)$ can be approximated as 
\begin{align}\label{MargLikEq}
 p(\obs) \approx \widehat{p}(\obs)=\dfrac{1}{JNT}
\sum_{j=1}^J  \sum_{t=1}^T \sum_{n=1}^N \beta_{t,j}^{(n)}.
\end{align}


\subsection{Approximation of the marginal posteriors}

An approximation of the marginal posterior distribution of $\bm{\theta}$ can be obtained 
\begin{align}\label{equ:margipostTheta}
 p({\bm \theta}|\obs) &= \int_{\Sigma} p(\bm{\theta}, {\bm \Sigma}|\obs) d{\bm \Sigma}\approx \widehat{p}(\bm{\theta}|\obs) = \sum_{t=1}^T \sum_{n=1}^N \bar{\alpha}_{t}^{(n)} \cdot  \delta({\bm \theta}-{\bm \theta}_t^{(n)}), 
\end{align}
where
\begin{align}
\bar{\alpha}_{t}^{(n)} =\frac{ \sum_{j=1}^J \beta_{t,j}^{(n)} }{\sum_{i=1}^J\sum_{v=1}^T \sum_{m=1}^N \beta_{v,i}^{(m)}}.
\end{align}
Moreover, we can assign a weight to each drawn matrix above ${\bm \Sigma}^{(j)}$, approximating the marginal posterior of the covariance matrix
\begin{align}\label{equ:sigma weights}
 p({\bm \Sigma}^{(j)}|\obs) &= \int_{{\bm \Theta}} p(\bm{\theta}, {\bm \Sigma}^{(j)}|\obs) d\bm{\theta} \approx \frac{ \sum_{t=1}^T \sum_{n=1}^N \beta_{t,j}^{(n)} }{\sum_{i=1}^J\sum_{v=1}^T \sum_{m=1}^N \beta_{v,i}^{(m)}} \delta\left({\bm \Sigma}-{\bm \Sigma}^{(j)}\right), \\
 &=\bar{\lambda}_j \cdot \delta\left({\bm \Sigma}-{\bm \Sigma}^{(j)}\right),
\end{align}
where 
\begin{align}
\bar{\lambda}_j=\frac{ \sum_{t=1}^T \sum_{n=1}^N \beta_{t,j}^{(n)} }{\sum_{i=1}^J\sum_{v=1}^T \sum_{m=1}^N \beta_{v,i}^{(m)}}.
\end{align}
Then, the marginal posterior of the covariance matrix is approximated as 
\begin{align}\label{equ:margipostSIGMA}
p({\bm \Sigma}|\obs) \approx \widehat{p}({\bm \Sigma}|\obs) =  \sum_{j=1}^J\bar{\lambda}_j \cdot \delta\left({\bm \Sigma}-{\bm \Sigma}^{(j)}\right).
\end{align}
 For instance, a minimum mean square error estimator of ${\bm \Sigma}$ can be approximated as 
$$
\widehat{{\bm \Sigma}}=\sum_{j=1}^J \bar{\lambda}_j {\bm \Sigma}^{(j)},
$$
and approximations of high-order moments $p({\bm \Sigma}|\obs)$ can also be  obtained. Table \ref{TablaApprox} summarizes all the weights and the corresponding distributions.

\begin{table}[h!]
		\centering
	\caption{Summary of the weights and the corresponding distributions.  \label{TablaApprox}}
	\begin{tabular}{|c|c||l|}
	\hline
	Distribution to approximate    & Normalized weights &  Additional information\\
	\hline
	\hline
	 &  &    \multirow{13}{*}{$\beta_{t,j}^{(n)}=\frac{\ell({\bf Y}|{\bm \theta}_t^{(n)}, {\bm \Sigma}^{(j)}) g_\theta({\bm \theta}_t^{(n)})}{q({\bm \theta}_t^{(n)}|{\bm \mu}_t,{\bm \Lambda}_t)} \cdot
  \frac{g_{\bm \Sigma}( {\bm \Sigma}^{(j)})}{ q_{\bm \Sigma}( {\bm \Sigma}^{(j)})}$}\\
 $p(\bm{\theta}|\obs,\widehat{{\bm \Sigma}}_\texttt{ML}^{(T)})$& $\widetilde{w}_{t}^{(n)}$ -- See Eqs. \eqref{FinalParApprox} and \eqref{aquiW}   & \\ 
 &   & \\
	$p(\bm{\theta}|\obs, {\bm \Sigma})$ & $\bar{\rho}_{t}^{(n)}({\bm \Sigma})=\frac{\ell({\bf Y}|{\bm \theta}_t^{(n)}, {\bm \Sigma}) g_\theta({\bm \theta}_t^{(n)})}{q({\bm \theta}_t^{(n)}|{\bm \mu}_t,{\bm \Lambda}_t)}$  &    \multirow{14}{*}{$\beta_{t,j}^{(n)}=\rho_{t}^{(n)}({\bm \Sigma}^{(j)})\cdot \gamma_j$}\\
	 &  & \\
	 $p(\bm{\theta},{\bm \Sigma} |\obs)$ &  $\bar{\beta}_{t,j}^{(n)}=\frac{\beta_{t,j}^{(n)}}{\sum_{i=1}^J  \sum_{v=1}^T \sum_{m=1}^N\beta_{v,i}^{(m)}}$ & \\
	 &  & \\
	 $p({\bm \theta}|\obs)$ & $\bar{\alpha}_{t}^{(n)} =\frac{ \sum_{j=1}^J \beta_{t,j}^{(n)} }{\sum_{i=1}^J\sum_{v=1}^T \sum_{m=1}^N \beta_{v,i}^{(m)}}$ & \\
	  & & \\
 $p({\bm \Sigma}|\obs)$ & $\bar{\lambda}_j=\frac{ \sum_{t=1}^T \sum_{n=1}^N \beta_{t,j}^{(n)} }{\sum_{i=1}^J\sum_{v=1}^T \sum_{m=1}^N \beta_{v,i}^{(m)}}$ & \\
  &  &\\
  $p(\obs)$ &  $\sum_{j=1}^J  \sum_{t=1}^T \sum_{n=1}^N \beta_{t,j}^{(n)}$ &\\ 
    & & \\
     \hline
	\end{tabular}

\end{table}

\subsection{Prior and proposal densities over covariance matrices}
\label{super-priors}

Consider a positive definite $K\times K$ matrix ${\bm \Sigma}$. The Wishart distribution  is defined on the space $\mathbb{R}^K\times\mathbb{R}^K$ of positive definite matrices. 
The corresponding pdf is 
\begin{equation}\label{WishartEQ}
g_{{\bm \Sigma}}({\bm \Sigma})=g_{{\bm \Sigma}}({\bm \Sigma}|{\bm \Phi},\nu)\propto |{\bm \Sigma}|^{\frac{\nu-K-1}{2}} \exp\left(-\frac{1}{2} \mbox{trace}({\bm \Phi}^{-1}{\bm \Sigma})\right), 
\end{equation}
where $|{\bm \Sigma}|$ denotes the determinant of the matrix ${\bm \Sigma}$, $\nu\geq K-1$ is the number of degrees of freedom and ${\bm \Phi}$ is a $K\times K$ {\it reference} covariance matrix.  It is possible to see $E_{g}[{\bm \Sigma}]=\nu {\bm \Phi} $. The Wishart distribution is often interpreted as a multivariate extension of the $\chi^2$ distribution.
\newline 
\newline
 \textbf{Choice of ${\bm \Phi}$ and $\nu$.}
We choose
\begin{align}
{\bm \Phi} =\frac{1}{\nu}\widehat{{\bm \Sigma}}_{\texttt{ML}}^{(T)}.
\end{align}
Recall that $\mathbb{E}_{g}[{\bm \Sigma}]=\nu {\bm \Phi} $.
In this sense, our approach is an {\it empirical Bayes scheme} since this parameter of the prior is chosen after looking at the data by ATAIS (see also data-based priors in \cite{Llorente22safePriors}). The parameter $\nu$ represents the degrees of freedom of the distribution. This value must be $\nu \geq K-1$, but for the generated matrices to be non-singular with probability 1 we need $\nu \geq K$.  For learning $\nu$, we can use again an  empirical Bayes approach maximizing the marginal likelihood $p(\obs)=p(\obs|\nu)$ in Eq. \eqref{MargLikEq}, i.e., we can find the $\nu^*$ such that $\nu^*=\arg\max p(\obs|\nu)$.
\newline 
\newline
  \textbf{Choice of the proposal pdf.} For simplicity, we assume  $q_{\bm \Sigma}( {\bm \Sigma})= g_{\bm \Sigma}( {\bm \Sigma})$, i.e., we choose a proposal density equal to the prior density. As a consequence, with this choice, we have $ \gamma_j=1$ in  \eqref{AquiGammaW}.  
\newline
\newline
{\bf Generation of random matrices according to a Wishart density.} When $\nu$ is an integer, the Wishart distribution represents the sums of squares (and cross-products) of $\nu$ vectors drawn from a multivariate Gaussian distribution. Specifically, given $\nu$ random vectors of dimension $K\times 1$, i.e.  ${\bf s}_i\sim \mathcal{N}({\bf 0}, {\bm \Phi})$, $i=1,\ldots, \nu$, the  generated matrix
$$
{\bm \Sigma}'=\sum_{i=1}^{\nu} {\bf s}_i{\bf s}_i^{\top},
$$  
%
%
is distributed as a Wishart density with $\nu$ degrees of freedom and  $K\times K$ scale matrix ${\bm \Phi}$. Then, we can employ the following sampling method:
\begin{enumerate}
\item Draw $\nu$ multivariate Gaussian samples ${\bf s}_i=[s_{i,1}, \ldots, s_{i,K}]^{\top} \sim \mathcal{N}({\bf 0}, {\bm \Phi})$, with $i=1,\ldots, \nu$.
\item Set ${\bm \Sigma}'=\sum_{i=1}^{\nu} {\bf s}_i{\bf s}_i^{\top}$.
\end{enumerate}

\section{Inverted layered importance sampling (ILIS)}\label{ILISsect}
The direct application  of Monte Carlo methods in the  complete space of $\bm{\theta}$ and $\bm{\Sigma}$  generally does  not provide good results. This  is the reason why we propose the ATAIS algorithm where the inference is carried out in two phases, first a set of weighted samples ${\bm \theta}_t^{(n)}$  and $\widehat{\bSigma}_\texttt{ML}$ are obtained. Then  the samples ${\bm \theta}_t^{(n)}$ are re-weighted and other weighted samples of matrices ${\bm \Sigma}^{(j)}$ are generated to perform a complete Bayesian inference.  
\newline
 In this section, we introduce an alternative method, conceptually simpler than ATAIS, called inverted layered importance sampling (ILIS), {based also on the division of the space, ${\bm \theta}$ and ${\bm \Sigma}$. ILIS can perform better than other Monte Carlo methods working in the  complete space,   but also presents some clear difficulties: this also emphasizes the relevance of ATAIS.}
See Table  \ref{InvLAISPepino} for a detailed description. ILIS starts by generating $N$ covariance matrices $\bm{\Sigma}^{(n)}$ and then it runs $N$ different (parallel and independent) MCMC chains with target density $p(\bm{\theta}|\bm{\Sigma}^{(n)}, \textbf{Y})$, i.e., the conditional posterior  of $\bm{\theta}$ given $\bm{\Sigma}^{(n)}$. { The underlying idea is to apply an IS scheme on the complete space $\{\bm{\theta},\bm{\Sigma}\}$, i.e., 
considering the complete posterior $p(\bm{\theta},{\bm \Sigma} |\obs)$ in Eq. \eqref{completePostEQ}, that is 
$$
 p(\bm{\theta},{\bm \Sigma} |\obs) = \frac{1}{p(\obs)}   \ell({\bf Y}|{\bm \theta}, {\bm \Sigma}) g_\theta({\bm \theta}) g_{{\bm \Sigma}}({\bm \Sigma}) 
 \propto \ell({\bf Y}|{\bm \theta}, {\bm \Sigma}) g_\theta({\bm \theta}) g_{{\bm \Sigma}}({\bm \Sigma}).
$$
Moreover, we assume to employ a proposal density of type $q(\bm{\theta},\bm{\Sigma})=q_\theta(\bm{\theta}|\bm{\Sigma})q_{{\bm \Sigma}}(\bm{\Sigma})$,
Recall that  $q$, $q_\theta$ and $q_{{\bm \Sigma}}$ must be normalized. The corresponding importance weight function in the full space is 
\begin{align}
w(\bm{\theta},\bm{\Sigma})&=\frac{ \ell({\bf Y}|{\bm \theta}, {\bm \Sigma}) g(\bm{\theta})g(\bm{\Sigma})}{q_\theta(\bm{\theta}|\bm{\Sigma})q_{{\bm \Sigma}}(\bm{\Sigma})}.
\end{align}
One good choice for $q_\theta$ would be
$$
q_\theta(\bm{\theta}|\bm{\Sigma})=  p(\bm{\theta}|\bm{\Sigma}, \textbf{Y}) = 
\frac{1}{Z({\bm \Sigma})}   \ell({\bf Y}|{\bm \theta}, {\bm \Sigma}) g_\theta({\bm \theta}) 
\propto \pi(\bm{\theta})= \ell({\bf Y}|{\bm \theta}, {\bm \Sigma}) g_\theta({\bm \theta}),
$$
where $Z(\bSigma)=\int _{\Theta}  \ell({\bf Y}|{\bm \theta}, {\bm \Sigma}) g_\theta({\bm \theta}) d{\bm \theta}$. This proposal is suggested and inspired by the success of the ATAIS approach. However, the choice above presents two issues:
\begin{itemize}
\item We need to be able to draw from. We can use an MCMC chain for generating samples distributed as $p(\bm{\theta}|\bm{\Sigma}, \textbf{Y})$, after a burn-in period \cite{Robert04,Liu04b,MHluca};
\item We need to be able to evaluate $Z(\bSigma)$, which is generally unknown. Thus, an estimation of this normalizing constant,  $\widehat{Z}({\bm \Sigma}) \approx Z({\bm \Sigma})$, is required \cite{Chib01,Llorente20}.
\end{itemize}
Indeed, in this case, the importance weight would be:
\begin{align}
w(\bm{\theta},\bm{\Sigma})&
=\frac{ \ell({\bf Y}|{\bm \theta}, {\bm \Sigma}) g(\bm{\theta})g(\bm{\Sigma})}{\frac{1}{Z({\bm \Sigma})}   \ell({\bf Y}|{\bm \theta}, {\bm \Sigma}) g_\theta({\bm \theta}) q_{{\bm \Sigma}}(\bm{\Sigma})}, \\
&= Z({\bm \Sigma})\frac{g(\bm{\Sigma})}{  q_{{\bm \Sigma}}(\bm{\Sigma})},
\end{align}
and the evaluation of  $Z(\bSigma)$ is required.} Other important considerations are remarked below:
\begin{itemize}
\item {For each $\bSigma^{(n)}\sim q_{\bSigma}(\bSigma)$, we need an MCMC chain to draw from $q_\theta(\bm{\theta}|\bm{\Sigma})=  p(\bm{\theta}|\bm{\Sigma}, \textbf{Y}) $.} Each MCMC algorithm produces a chain of $T$ vectors, i.e., ${\bm \theta}_1^{(n)},....{\bm \theta}_T^{(n)}$. All these vector are weighted with the same weight {$\gamma_n=Z(\bSigma^{(n)})\dfrac{ g_{\bSigma}(\bSigma^{(n)})}{q_{\bSigma}(\bSigma^{(n)})}$}.
\item ILIS can be seen as a Monte Carlo scheme that combines IS and MCMC techniques, based on {\it two layers} (a global IS scheme that employs internally MCMC chains). With respect to the techniques in \cite{LAIS17,Llorente22mcmcDriven}, we can interpret that the IS part forms the upper layer (deciding the parameter $\bSigma^{(n)}$ to be used in $q_\theta(\bm{\theta}|\bm{\Sigma}^{(n)})=  p(\bm{\theta}|\bm{\Sigma}^{(n)}, \textbf{Y}) $), whereas the MCMC chains are generated in the lower layer of ILIS with target/invariant densities $p(\bm{\theta}|\bm{\Sigma}^{(n)}, \textbf{Y})$ (note that this pdf plays the role of a proposal $q_\theta$ in the global IS scheme). All the states of one chain are weighted with the weight {$\gamma_n=Z(\bSigma^{(n)})\dfrac{ g_{\bSigma}(\bSigma^{(n)})}{q_{\bSigma}(\bSigma^{(n)})}$  (i.e., the layers are switched with respect to \cite{LAIS17,Llorente22mcmcDriven})}.
\item Note that again, as in ATAIS, we have a different number of samples with respect to ${\bm \theta}$ (i.e., $NT$), and  with respect to $\bSigma$ (i.e., $N$), which recalls the recycling Gibbs scheme in \cite{RecGibbs}.
\end{itemize}
The complete posterior approximation by ILIS is given by
\begin{align}
 \widehat{p}(\bm{\theta}, {\bm \Sigma}|\obs)=\frac{1}{T}\sum_{t=1}^T \sum_{n=1}^N \bar{\gamma}_n \cdot \delta({\bm \theta}-{\bm \theta}_t^{(n)}) \delta\left({\bm \Sigma}-{\bm \Sigma}^{(n)}\right), \quad \mbox{ with } \quad \bar{\gamma}_{n}=\frac{\gamma_{n}}{\sum_{i=1}^N\gamma_{i}}.
\end{align}
Even if ILIS seems at least theoretically simpler than ATAIS, {we need to compute the approximations of $Z(\bSigma^{(n)})$} and its performance is not particularly good compared to ATAIS, as we show in the numerical simulations. { The main reason is that ATAIS, in the first part of the inference, finds the regions of high probability and also uses this information to tune the proposal $q_{\bSigma}(\bSigma)$ for the matrix generation, in the second part.}
\newline
\newline
{\bf Joint use of ATAIS and ILIS.} One interested practitioner  could use the first part of ATAIS in Table \ref{AIS_AutoTemp} ``to feed'' with good parameters the proposal density $q_{\bSigma}(\bSigma)$ and the proposals used inside the $N$ parallel MCMC chains (i.e., design good proposals for ILIS). However,  with respect to the complete ATAIS, this scheme has a higher computational cost in terms of generated samples and likelihood evaluation. Indeed, recall that the second part of ATAIS does not require any additional sample generation and likelihood evaluation.

\begin{table}[h!]
	\caption{Inverted layered importance sampling (ILIS) \label{InvLAISPepino}}
	\begin{tabular}{|p{0.95\columnwidth}|}
		\hline
		\begin{enumerate}
			\item Choose $N$, $T$, $\bm{\mu}_0$ and $q_{\bSigma}(\bSigma)$ as well as the the proposal densities and structure of $M$ possibly different MCMC methods.
			\item For $j=1,\dots, J$:
			\begin{enumerate}
				\item Generate $\bSigma^{(j)}\sim q_{\bSigma}(\bSigma)$.
				\item Generate $N$ different MCMC chains of length $T$ obtaining $\{\bm{\theta}_t^{(n)}\}_{t=1}^T$, with the conditional distribution
				$$
				\bar{\pi}(\bm{\theta}|\bm{\Sigma}^{(j)}, \textbf{Y}) \propto   \ell({\bf Y}|{\bm \theta}, {\bm \Sigma}^{(j)}) g_\theta({\bm \theta}),
				$$ 
				as a target density.
				\item {Approximate $Z(\bSigma^{(j)})$, using some procedure for estimation normalizing constants \cite{Chib01,Llorente20}, i.e., obtain $\widehat{Z}(\bSigma^{(j)}) \approx Z(\bSigma^{(j)})=\int _{\Theta}  \ell({\bf Y}|{\bm \theta}, {\bm \Sigma}^{(j)}) g_\theta({\bm \theta}) d{\bm \theta}$.}
				\item Assign to each pair $\{\btheta_t^{(n)}, \bSigma^{(n)}\}$ the weight $\gamma_j=\widehat{Z}(\bSigma^{(j)})\dfrac{ g_{\bSigma}(\bSigma^{(j)})}{q_{\bSigma}(\bSigma^{(j)})}$, for all $j=1,...,J$ and $t=1,...,T$.
			\end{enumerate}
		\end{enumerate} \\
		\hline 
	\end{tabular}
\end{table}

\section{Simulations}
\label{sec:Simul}

{ We test the proposed scheme in numerous different numerical examples, comparing it with different benchmark schemes.  We have also tested several possible scenarios and variants, such as different noise perturbations in the observation model and the use of mini-batches. The dimension of the space is:
\begin{itemize}
\item $D=8$ for the model in Section \ref{LocSect},    
\item $D=12$ for the model in Section \ref{SectMO},
\item  $D=7$ for the model in Section \ref{BioExample},
 \item and $D=59$ for the model in Section \ref{GraphExamples}.
\end{itemize}
We recall that, when a Monte Carlo method is tested, we should know some groundtruths of the target density, i.e., some true features of the posterior distribution (such as modes, expected values, variances, and any other moments). However, they are generally unknown even in experiments with artificial data. Hence, in many cases, the only possibility for the authors/practitioners is to consider as groundtruths the true values used for generating the data, i.e., ${{\bm \theta}}_{\texttt{true}}$ and ${\bm \Sigma}_{\texttt{true}}$ here. Clearly, if the number of data is not enough ${{\bm \theta}}_{\texttt{true}}$ and ${\bm \Sigma}_{\texttt{true}}$ can be quite different from the true groundtruth values (i.e., features of the posterior).
 When the example is low-dimensional, an alternative is to employ an expensive grid (or another determinist approach) for approximating the true features of the posterior  (modes, expected values etc.), and use those approximated values as groundtruths.  
 In this work, for ${\bm \Sigma}$, we have another possible groundtruth for the ML estimation which is the following:
\begin{align}\label{MatrixTRUEeq2}
{\bm \Sigma}_{\texttt{ML}}\approx \frac{1}{R} \sum_{r=1}^R\left({\bf y}_r- {\bf f}_r({{\bm \theta}}_{\texttt{true}})\right)\left({\bf y}_r- {\bf f}_r({{\bm \theta}}_{\texttt{true}})\right)^{\top},
\end{align}
where ${\bm \theta}_{\texttt{true}}$ are the true values used for generating the data, according to the model considered in the specific example (recall that we do not know the true values of ${\bm \theta}_{\texttt{MAP}}$). Clearly, as stated before, ${\bm \Sigma}_{\texttt{true}}$ can be also employed as groundtruth, and the difference between ${\bm \Sigma}_{\texttt{ML}}$ and ${\bm \Sigma}_{\texttt{true}}$ is minimal, if we are analyzing enough data. 
\newline 
We average all the results over different independent runs. At each run, the initializations of the ATAIS parameters (and other schemes) are chosen randomly around a vector of zeros for ${\bm \theta}$ and an identity matrix for ${\bm \Sigma}$ (with a standard deviation of $4$; for the covariance matrix ${\bm \Sigma}$, we always consider an initial diagonal matrix with the same random element in the diagonal).  
}

\subsection{Localization in wireless sensor network {- dimension of the space $D=8$}}\label{LocSect} 
{\color{black}
	In this experiment, we aim to determine the location of a target based on wireless sensor measurements. We can represent the target position as a random vector $\bm{\theta} \in \mathbb{R}^2$. 
	We have a wireless network of $K=3$ sensors, whose positions are known and labeled as ${\bf s}_1, . . . , {\bf s}_K$. We collect $R$ measurements from each sensor, and these measurements follow a certain distribution. Let's recall that each observation has the following form:
	\begin{equation}\label{key}
		y_r = \textbf{f}_r(\bm{\theta}) + \textbf{v}_r,
	\end{equation}
	with $\textbf{f}_r:\mathbb{R}^2\to \mathbb{R}^3$ given by:
	\begin{equation}\label{key}
		\textbf{f}_r(\bm{\theta}) = \Big[-A\log(\parallel\bm{\theta} - {\bf s}_1\parallel^2), \, -A\log(\parallel\bm{\theta} - {\bf s}_2\parallel^2), \, -A\log(\parallel\bm{\theta} - {\bf s}_3\parallel^2)\Big],
	\end{equation}
	%
	i.e., 
	$$
	f_{r,i}(\bm{\theta})=-A\log(\parallel\bm{\theta} - {\bf s}_i\parallel^2),
	$$
	for $i=1,2,3$.
	The error term is  $\bf{v}_r\sim \mathcal{N}(\bm{0}, {\bm \Sigma}_{\texttt{true}})$, with ${\bm \Sigma}_{\texttt{true}}\in\mathbb{R}^{3\times 3}$ being a diagonal matrix with diagonal elements 1, 2 and 3.
	The parameter $A$ is a constant that determines the rate at which the signal strength decreases with distance and is fixed at 10. This value can be influenced by various factors, such as environmental conditions or manufacturing processes. The values of variance of the sensors, as stated before, is unknown for each sensor.\\
	We consider a scenario with $K=3$ sensors, which makes the complete dimension of the problem to be
	$$
	D=  M + \dfrac{K(K+1)}{2} = 2 + \dfrac{3(3+1)}{2}=8.
	$$ 
	The positions of these sensors are given by: ${\bf s}_1 = [0.5, 1]$, ${\bf s}_1 = [3.5, 1]$ and ${\bf s}_3 = [2, 3]$. The positions of the target (and parameter we want to estimate) is $\bm{\theta}_{\texttt{true}}=[2.5,\,2]$. 
	In this scenario, 50 observation vectors were generated. For comparison purposes, the prior over $\bm{\theta}$ was set as uniform, i.e., $g(\bm{\theta}) \propto 1$. 
\newline
\newline		%
{\bf Testing ATAIS.} In ATAIS, we set $T=40$ iterations and $N=50$ particles per iteration. We consider a  Gaussian proposal density for the ${\bm \theta}$-space with initial mean $[0,0]^{\top}$, with a diagonal initial covariance matrix of $6  {\bf I}_2$. The initial covariance matrix  $\widehat{\bSigma}_{\texttt{ML}}^{(0)}$  is set to the identity matrix $\textbf{I}_3$. 
 All the results are averaged over $10^3$ independent runs.
 First, in Figure \ref{MERDA_DI_nomi_por_Ernesto}, we provide the mean absolute error (MAE) versus $N$ with $T=50$ and $T=100$ (two curves), obtained by ATAIS in estimating in the estimation of ${\bm \theta}_{\texttt{MAP}}$  (using ${\bm \theta}_{\texttt{true}}$ as the groundtruth). The error decreases as we increase the number of samples for both values of  $T=50$ and $T=100$. { In Figure \ref{MERDA_DI_nomi_por_Ernesto2}, we give the MAE versus $T$ with $N=50$ and $N=100$. Again, we obtain the expected behavior: the error decreases as $T$ grows and/or $N$ increases.}  Additionally, other MAE values  (also in estimating the covariance matrix ${\bm \Sigma}$) obtained by ATAIS are given in Table \ref{tab:mae-location-T-N}.
 {Recall that we consider  $\btheta_{\texttt{true}}$ and  
$\bSigma_\texttt{ML}$ - computed as in Eq \eqref{MatrixTRUEeq2} - as the groundtruths of this experiment. In this table, we show the two partial  MAEs for $\btheta_{\texttt{true}}$ and  
$\bSigma_\texttt{ML}$, and the MAE in the complete space, as well. Figure \ref{fig:ex-1-path} illustrates the evolution  of the components of $\widehat{\btheta}_{\texttt{MAP}}$ and $\widehat{\bSigma}_{\texttt{ML}}$ with the iterations, in one run of ATAIS.}
\newline
\newline
{ {\bf Comparison with ILIS and a ``helped'' MH. }  We compare the ATAIS results  with (a) ILIS using Metropolis Hastings (MH) chains and (b) a unique MH chain working only in the ${\bm \theta}$-space, keeping fixed the covariance matrix to the maximum likelihood estimation $\widehat{\Sigma}_{\texttt{ML}}^{(T)}$ obtained by ATAIS in its first part. Namely, this MH is helped by using and keeping $\widehat{\Sigma}_{\texttt{ML}}^{(T)}$ given by ATAIS and works only in the ${\bm \theta}$-space.} 
 For ILIS, we consider as prior and proposal density for the covariance  matrices a Wishart distribution with $\nu=4$ degrees of freedom and a reference matrix $\Phi=3\textbf{I}_3$. Then, for ${\bm \theta}$-space, we use MH chains with random walk Gaussian proposal density (starting in $[0, 0]$ and diagonal covariance $0.05 \bf{I}_2$) and with length $T=200$. We set $J=10$, i.e., we generate $J=10$ possible matrices and we have $J=10$ parallel chains considering different target densities (i.e., different conditional posterior pdfs, each one considering a different covariance matrix).
Finally, for the unique/single MH chain addressing the conditional posterior keeping fixed $\widehat{\Sigma}_{\texttt{ML}}^{(T)}$ (obtained by ATAIS),  we consider again a random walk Gaussian proposal density (with initial $[0, 0]^{\top}$ and diagonal covariance $0.05 \bf{I}_2$) and with length $T=2000$.
\newline
Note that the number of evaluations of the non-linear model ${\bf f}$ does not differ in the different methods:
 in ATAIS is only $NT=2000$, in ILIS, is $JT=2000$, whereas in the single MH  we have  $T=2000$ evaluations of ${\bf f}$. Therefore, all techniques have the same evaluations of the non-linear model ${\bf f}({\bm \theta})$ that corresponds to the main cost in the likelihood evaluation.
 \newline	
 Our goal here is to approximate the MAP estimation of ${\bm \theta}$ by all these schemes.
 The results are given in Figure \ref{fig:atais-ilis}. We can see the final ATAIS approximates of  MAP of ${\bm \theta}$ (using ${\bm \theta}_{\texttt{true}}$ as the groundtruth), represented with green squares, whereas  the estimations of ILIS are presented with red circles.  The results provided by the single MH chain are depicted with blue diamonds in Figure \ref{fig:atais-ilis}. Looking at Figure \ref{fig:atais-ilis}, it is clear that ATAIS (green squares) provides the best performance estimating $\btheta_{\texttt{true}}$ (black cross) better than ILIS (red circles). In addition, we can see that the single MH chain using the covariance matrix estimated by ATAIS (blue diamonds) shows better results in most of the cases than ILIS, but they are still worse than ATAIS.
\newline
Note that, in this example, the posterior is very narrow, which can make it hard for the generation of  samples in regions with high posterior evaluation, this is why the adaptation of the proposal scale matrix in step 2d of Table \ref{AIS_AutoTemp}  is quite important.  To improve even more the exploration, at some iteration the scale matrix of the proposal densities can be periodically increased to allow exploration of areas away from the narrow mode {as described in Section \ref{AdapCOVprop}.}

 	\begin{figure}[H]
	\centerline{
		\subfigure[MAE versus $N$\label{MERDA_DI_nomi_por_Ernesto}]{\includegraphics[width=8cm]{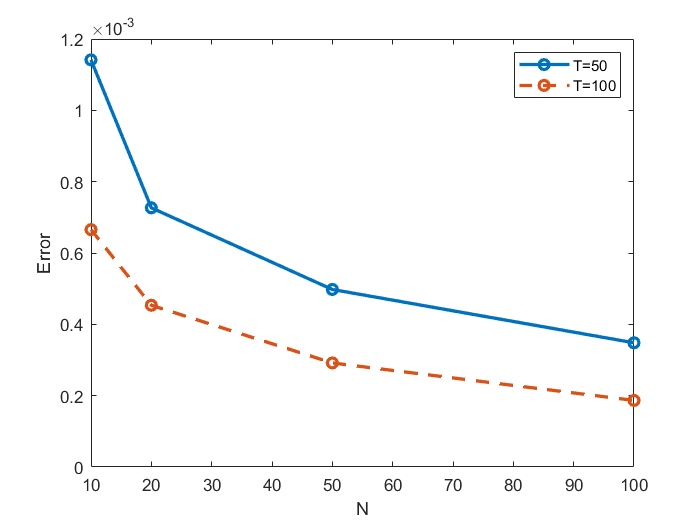}}
		\subfigure[{MAE versus $T$} \label{MERDA_DI_nomi_por_Ernesto2}]{\includegraphics[width=8.2cm]{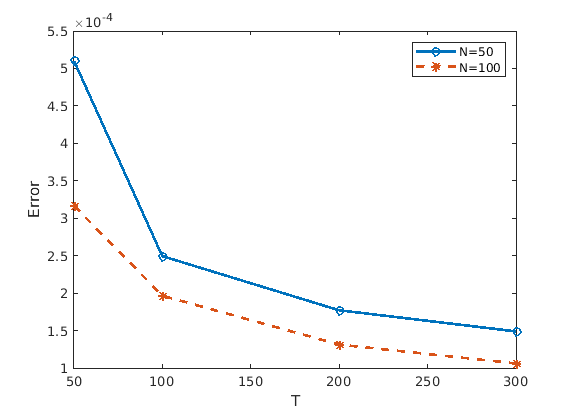}}
		}
		\caption{Location Example. MAE in the estimation of ${\bm \theta}_{\texttt{MAP}}$  (using ${\bm \theta}_{\texttt{true}}$ as the groundtruth), with different number of particles by ATAIS versus {\bf (a)} $N$ { with fixed $T$ and with {\bf (b)} $T$ with fixed $N$}.   }
		\label{fig:error media ambos}
	\end{figure}
	
\begin{figure}[H]
	\subfigure[{ $\widehat{{\bm \theta}}_{\texttt{MAP}}^{(t)}$}]{\includegraphics[width=8cm]{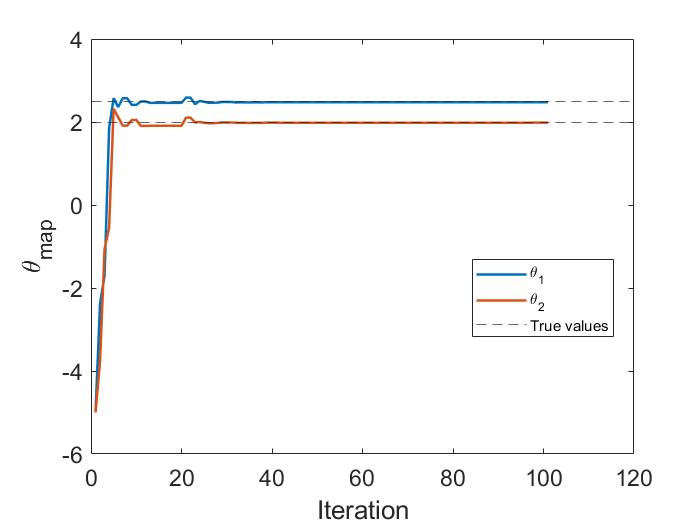}}
	\subfigure[{ $\widehat{{\bm \Sigma}}_{\texttt{ML}}^{(t)}$}]{\includegraphics[width=8cm]{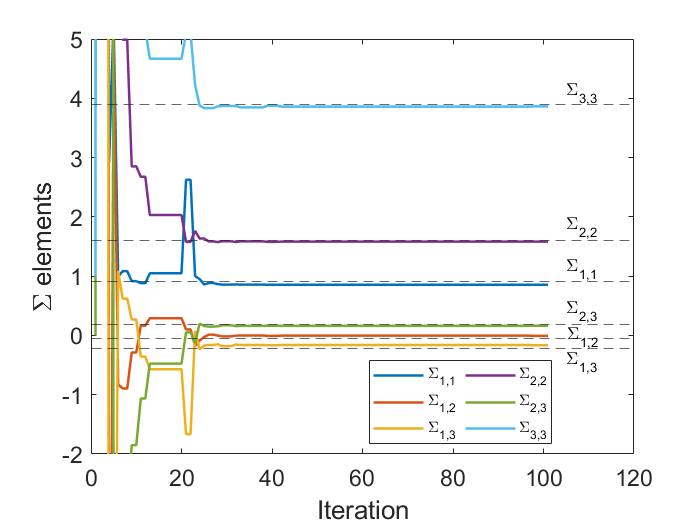}}
	\caption{{Location example. Convergence of the components in $\widehat{{\bm \theta}}_{\texttt{MAP}}^{(t)}$ and $\widehat{{\bm \Sigma}}_{\texttt{ML}}^{(t)}$ in one run of ATAIS algorithm.}}
	\label{fig:ex-1-path}
\end{figure}

\noindent
\textbf{Credible interval with $95\%$ of probability for the matrix}. In order to show how to perform a complete Bayesian inference over the covariance matrix ${\bm \Sigma}$ as well, we consider a Wishart  proposal with $\nu=100$ (degrees of freedom) with a reference matrix ($\bm{\Phi}$) equal to $\widehat{\bm{\Sigma}}_{\texttt{ML}}^{(T)}$ (i.e., we apply the second part of ATAIS). With this proposal distribution, we generate $J=1000$ matrices and assign a weight to each of them following Eq. \eqref{equ:sigma weights}. Applying resampling (exactly $J$ times) according to the normalized weights $\{\bar{\lambda}_j\}_{j=1}^J$, we calculate the percentiles 0.025 and 0.975 for each component to get a credible interval for the covariance matrix $\bm{\Sigma}$, as shown below (where we have averaged over $100$ independent runs) in Eq \eqref{equ:interval_location}. The first part of ATAIS was performed with $T=50$ and $N=50$, obtaining a confidence interval with $\alpha=0.05$ for each value of the matrix,
\begin{equation}\label{equ:interval_location}
	\begin{pmatrix}
		[ 0.6697, 1.3594] &  [-0.3206, 0.2958] &  [-0.6946, 0.3115] \\
		[-0.3206, 0.2958] &  [ 1.2584, 2.4827] &  [-0.4709, 0.8096] \\
		[-0.6946, 0.3115] &  [-0.4709, 0.8096] &  [ 2.9975, 5.8437] 
	\end{pmatrix}
\end{equation}
Note that the  covariance matrix ${\bm \Sigma}$ represents the covariance among the different sensors in the network.
In Figure \ref{fig:hist_location}, we can see the histograms obtained by the resampled particles of each component of $\bSigma$ (after performing resampling according to the weights $\{\bar{\lambda}_j\}_{j=1}^{J}$). We must remark on how the histograms corresponding to the null components of $\bSigma_{\texttt{true}}$ have the mode very close to the value 0.

\begin{figure}[h]
	\centering
	\includegraphics[scale=.6]{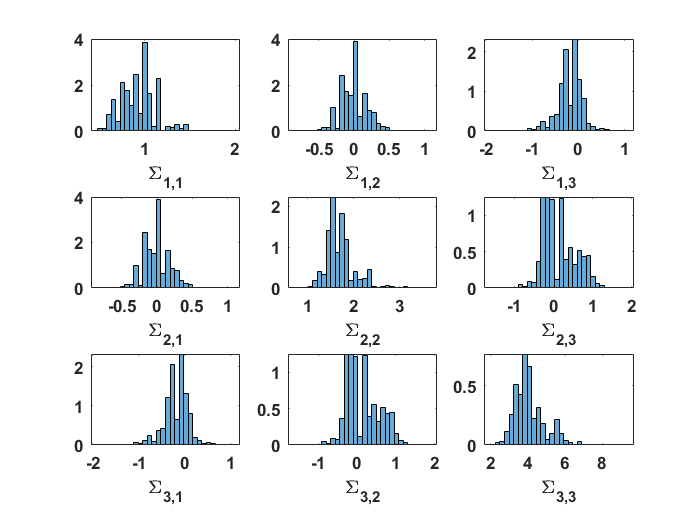}
	\caption{ Location example: histogram of the components of ${\bm \Sigma}$, denoted as $[{\bm \Sigma}]_{i,j}=\bSigma_{i,j}$, of the covariance matrices after resampling according to the weights $\bar{\lambda}_j$, for the location example.} 
	\label{fig:hist_location}
\end{figure}


\begin{figure}[H]
	\centering
	\includegraphics[width=7.3cm]{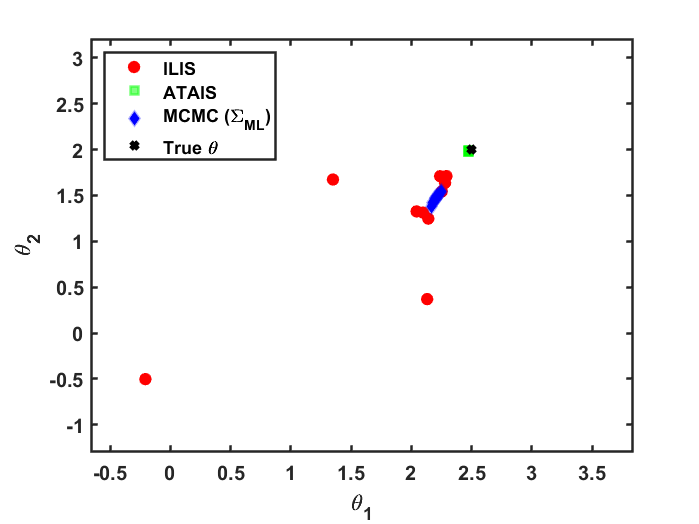}
	\caption{ Location example:   Estimations of $\btheta_{\texttt{MAP}}$ with different schemes. the green squares represent the estimation with ATAIS; the red circles depict for the estimations of ILIS with a covariance matrix generated from a Wishart distribution. The blue diamonds represent the estimations of the single MH addressing a conditional posterior as invariant density,  where ${\bm \Sigma}$ is fixed to the value of estimation $\bm{\Sigma}_{\texttt{ML}}$ obtained by ATAIS. The black cross shows ${\bm \theta}_{\texttt{true}}$.}
	\label{fig:atais-ilis}
\end{figure}

 \begin{table}[h!]
 	\centering
 	\caption{{Localization example: MAE averaged over $100$ simulations of ATAIS for estimating the $\bm{\theta}_{\texttt{true}}$, $\bSigma_\texttt{ML}$. Recall that its groundtruth is computed as in Eq. \eqref{MatrixTRUEeq2}. The complete MAE in the whole space is also given.}  }
 	\label{tab:mae-location-T-N}
 	
 	\begin{tabular}{l|lll|l|lll|}
 		\cline{2-4} \cline{6-8}
 		& \multicolumn{3}{l|}{{\bf Varying in $N$ ($T=50$)}}                           &    & \multicolumn{3}{l|}{{\bf Varying in $T$ $(N=100)$}}                          \\ \hline
 		\multicolumn{1}{|l|}{$N$}   & \multicolumn{1}{l|}{$\btheta_{\texttt{true}}$} & \multicolumn{1}{l|}{$\bSigma_{\texttt{ML}}$} & {\bf Complete MAE}   & $T$  & \multicolumn{1}{l|}{$\btheta_{\texttt{true}}$} & \multicolumn{1}{l|}{$\bSigma_{\texttt{ML}}$} & {\bf Complete MAE}  \\ \hline
 		\multicolumn{1}{|l|}{5}   & \multicolumn{1}{l|}{0.0377} & \multicolumn{1}{l|}{0.8934} & 0.7378  & 5  & \multicolumn{1}{l|}{0.1740} & \multicolumn{1}{l|}{2.4068} & 2.0009 \\ \hline
 		\multicolumn{1}{|l|}{12}  & \multicolumn{1}{l|}{0.0207} & \multicolumn{1}{l|}{0.0443} & 0.0400 & 10 & \multicolumn{1}{l|}{0.0758} & \multicolumn{1}{l|}{0.4292} & 0.4834   \\ \hline
 		\multicolumn{1}{|l|}{25}  & \multicolumn{1}{l|}{0.0205} & \multicolumn{1}{l|}{0.0443} & 0.0399 & 20 & \multicolumn{1}{l|}{0.0328} & \multicolumn{1}{l|}{0.3002} & 0.3649 \\ \hline
 		\multicolumn{1}{|l|}{50}  & \multicolumn{1}{l|}{0.0205} & \multicolumn{1}{l|}{0.0442} & 0.0399 & 30 & \multicolumn{1}{l|}{0.0205} & \multicolumn{1}{l|}{0.0441} & 0.0398 \\ \hline
 		\multicolumn{1}{|l|}{100} & \multicolumn{1}{l|}{0.0204} & \multicolumn{1}{l|}{0.0440} & 0.0400 & 50 & \multicolumn{1}{l|}{0.0204} & \multicolumn{1}{l|}{0.0440} & 0.0400 \\ \hline
 	\end{tabular}
 \end{table}

 \subsection{Multi-output model - { dimension of the space $D=12$}}\label{SectMO}
	In this second experiment, we consider the following isotopic multi-output model, using the notation in Eq. \eqref{NotationAqui},  
	\begin{equation}\label{key}
		\textbf{y} = \textbf{f}(\bm{\theta},\tau) + \textbf{v},
	\end{equation}
	where the vector function $\textbf{f}(\bm{\theta},\tau):\mathbb{R}^2\times \mathbb{R} \to \mathbb{R}^4$ with $\bm{\theta}=[\theta_1,\theta_2]^{\top}$ is given by the components
	\begin{align}\label{key}		
			f_{1}(\bm{\theta},\tau) &=f_{1}(\theta_1,\tau)= \theta_1 \sin(\tau) \tau, \nonumber \\
			f_{2}(\bm{\theta},\tau) &=f_{2}(\theta_2,\tau)= \theta_2 \cos(\tau) \tau^2,  \\
			f_{3}(\bm{\theta},\tau) &=  (\theta_1 + \theta_2) \sin(\tau) \cos(\tau), \nonumber \\
			f_{4}(\bm{\theta},\tau) &=f_{4}(\theta_2,\tau)=  \theta_2 \tau^2. \nonumber		
	\end{align}
	The error term ${\bf v} \sim \mathcal{N}(\textbf{0}, \bm{\Sigma}_{\texttt{true}})$ with 
	\[ \bm{\Sigma}_{\texttt{true}} = 
	\begin{pmatrix}
		0.1 & 0.3 & 0.16 & 0\\
		0.3 & 1.05 & 0 & 0 \\
		0.16 & 0 & 2 & 0 \\
		0 & 0 & 0 & 2.95
	\end{pmatrix}.
	 \]
	The true value of the vector ${\bm \theta}$ is set to $\bm{\theta}_{\texttt{true}}=[0.2, 0.1]^\top$. In this case de dimension of the observations remains at $K=4$, with $\bm{\theta}$ of dimension $M=2$, which makes a total dimension of inference space of
	$$
	D=  M + \dfrac{K(K+1)}{2} = 2 + \dfrac{4(4+1)}{2}=12.
	$$ 
{\bf Remark.} This example shows clearly the difficulty of performing an estimation of $\bm{\Sigma}$ directly from the vectors observed, $\{\textbf{y}_r\}_{r=1}^R$. This is due to the vectors not sharing the same mean (i.e., the functions $f_i$ which also depends on $\tau$).
\newline
\newline
 The prior for $\bm{\theta}$ is assumed to be uniform and improper, i.e., $g(\bm{\theta})\propto 1$. 
	As in the previous example, we also test three algorithms: (a) complete ATAIS, (b) ILIS using  Metropolis-Hastings (MH) chains and (c) a single MH chain working exclusively in the $ \btheta $-space, maintaining fixed the covariance matrix and equal to $\widehat{\Sigma}_{\texttt{ML}}^{(T)}$ given by ATAIS, as in the previous experiment.  We aim to approximate the MAP estimate of $ \btheta$. All the results are averaged over 1000 independent runs.
\newline	
{\bf ATAIS results.} We employ in ATAIS the same specifications as in the previous example, i.e., $N=50, T=40$. We use a Gaussian proposal with the initial mean at $[0, 0]$ and the initial covariance matrix is $6 \bf{I}_2$. The initial covariance matrix for the observations, $\widehat{\bSigma}_{\texttt{ML}}^{(0)}$, is the identity matrix, $\textbf{I}_4$.
{Figure \ref{fig:ex-3-path} depicts the evolution of the estimators $\widehat{{\bm \theta}}_{\texttt{MAP}}^{(t)}$ and $\widehat{{\bm \Sigma}}_{\texttt{ML}}^{(t)}$ versus iteration $t$, in one run of ATAIS algorithm.}
{In Figures \ref{ME_CAGO_EN_TODOS_A} and  \ref{ME_CAGO_EN_TODOS_B}, we also provide the MAE versus the number of particles $N$ and $T$, obtained by ATAIS in estimating $\btheta_{\texttt{MAP}}$, using $ \btheta_{\texttt{true}}$ as groundtruth.}  
\newline	
	Table \ref{tab:mae-multioutput-T-N} provides other MAE values also in estimating the covariance matrix $ \bSigma $, obtained by ATAIS. 
	As expected, the best results are obtained increasing the
	number of particles and the iterations, allowing a better exploration of the parameter space. Even if the error of the estimation of $\bm{\Sigma}_{\texttt{ML}}$ is quite small. This numerical example shows the strength of ATAIS, since in this multi-output problem the covariance matrix $\bm{\Sigma}$ cannot be approximately in advance directly from the data.
{ Additionally, considering the MAE in the whole state space, we also perform different simulations where the product $NT$ remains fixed, while the value of $T$ changes in $\{100, 200, 500, 1000, 2000\}$, so that  $N$  changes among these values  $\{200, 100, 40, 20, 10\}$. The results obtained are given in Figure \ref{Tu_puta_madre}.
We can observe the robustness of ATAIS: we obtain very small MAE values in all cases.
 }


\begin{figure}[H]
	\subfigure[{$\widehat{{\bm \theta}}_{\texttt{MAP}}^{(t)}$} \label{fig:ex-3-pathA} ]{\includegraphics[width=0.48\textwidth]{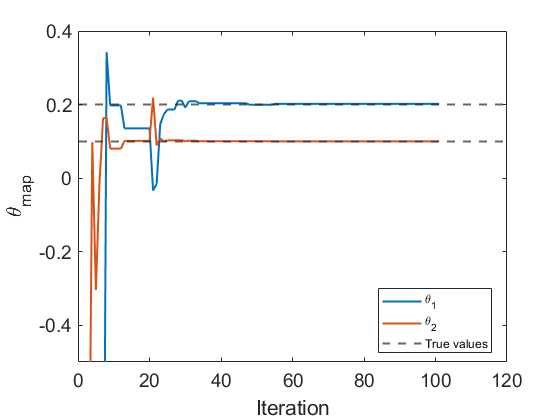}}
	\subfigure[{$\widehat{{\bm \Sigma}}_{\texttt{ML}}^{(t)}$} \label{fig:ex-3-pathB} ]{\includegraphics[width=0.48\textwidth]{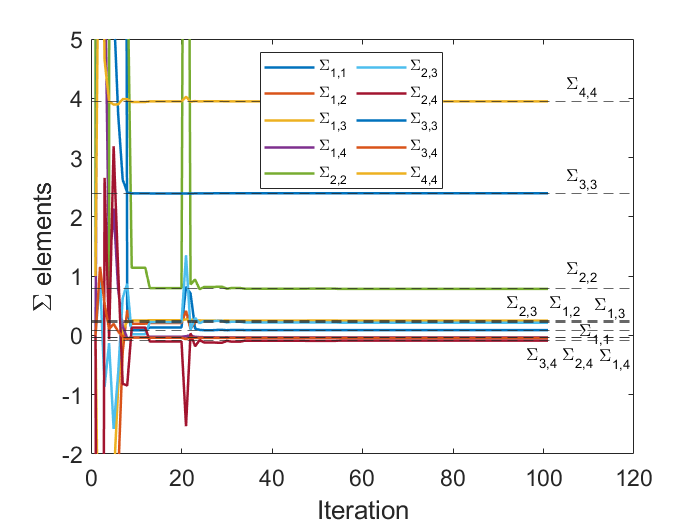}}
	\caption{{Multi-output example. Convergence of the components in the estimators $\widehat{{\bm \theta}}_{\texttt{MAP}}^{(t)}$ and $\widehat{{\bm \Sigma}}_{\texttt{ML}}^{(t)}$ in one run of ATAIS algorithm.}}
	\label{fig:ex-3-path}
\end{figure}

\noindent
{ {\bf Comparison with ILIS and an ``helped''-MH. }  We compare again the ATAIS results  with (a) ILIS using Metropolis Hastings (MH) chains and (b) a unique MH chain working only in the ${\bm \theta}$-space, keeping fixed the covariance matrix to the maximum likelihood estimation $\widehat{\Sigma}_{\texttt{ML}}^{(T)}$ obtained by ATAIS in its first part. Namely, this MH is helped by using and keeping $\widehat{\Sigma}_{\texttt{ML}}^{(T)}$ given by ATAIS, and works only n the ${\bm \theta}$-space.}  For ILIS, we consider as prior and proposal density for the covariance matrix a Wishart distribution with $\nu=5$ degrees of freedom and a reference matrix $\Phi=\frac{1}{\nu}\textbf{I}_4$. For working in the $ \btheta $-space we use MH chains with a  Gaussian random walk proposal density with initial mean [0,0] and diagonal covariance matrix $ 0.005\textbf{I}_2$. The length of the chains is $ T=200$. We generate $J=10$ possible matrices, thus we have $J=10$ parallel chains considering different target densities (each target considers a different covariance matrix). We consider that  all the chains provide a unique estimation of the MAP of $ \btheta$.
Finally, for the single MH chain addressing the conditional posterior fixing $ \widehat{{\bm \Sigma}}_{\texttt{ML}}^{(T)} $ (obtained by ATAIS), we consider a Gaussian random walk proposal density. This proposal density has an initial mean $[0,0]$ and has a diagonal covariance matrix $ 0.005\textbf{I}_2 $. The length of the chain is $ T=2000 $.
\newline		
In Figure \ref{Tu_puta_madre2}, we can see the final ATAIS estimates of the MAP of $ \btheta $ represented by green squares, while red circles represent the estimations of ILIS.
		The results provided by a single MH chain are displayed using blue diamonds. Looking at Figure \ref{Tu_puta_madre2}, it is clear that the ATAIS gives the best estimations of $\btheta_{\texttt{true}}$ (black cross) than the estimations of ILIS. Once again, we see how using the covariance matrix estimated by ATAIS (blue diamonds) gives better results than ILIS. In addition, we can see that the single MH chain using the covariance matrix estimated by ATAIS (blue diamonds) shows better results than ILIS in most of the cases, and some of them are comparable to ATAIS. Recall that, this third method has the advantage of using exactly $ \widehat{{\bm \Sigma}}_{\texttt{ML}}^{(T)} $ that is the estimation provided by ATAIS. Then, the success of this third method is mainly due to an ATAIS ability.
\newline	
\newline
\newline
	\textbf{Credible interval with $95\%$ of probability for the matrix}. In order to perform a complete Bayesian inference over the covariance matrix $\bm{\Sigma}$, we apply the second part of ATAIS. We consider a Wishart density as proposal (and prior) with $\nu=100$ (degrees of freedom) with a reference matrix ($\bm{\Phi}$) equal to $\widehat{\bm{\Sigma}}_{\texttt{ML}}^{(T)}$ (i.e., we apply the second part of ATAIS). With this proposal distribution, we generate $J=1000$ matrices and assign a weight to each of the following Eq. \eqref{equ:sigma weights}. Applying resampling according to the normalized weights $\{\bar{\lambda}_j\}_{j=1}^J$, we calculate the percentiles 0.025 and 0.975 for each component to get a credible interval for the covariance matrix $\bm{\Sigma}$, as shown below  in Eq \eqref{equ:intervalo_multi_N_50} (where we have averaged over $100$ independent runs). The first part of ATAIS was performed with $ T=50 $ and $ N=50 $.
		
	\begin{equation}\label{equ:intervalo_multi_N_50}
		\begin{pmatrix}
			[0.0614 ,  0.1586]  & [0.1669, 0.4483 ] & [ 0.0703, 0.4736] & [ -0.2578, 0.1985] \\ 
			[0.1669 ,  0.4483]  & [0.6125, 1.5325 ] & [-0.4739, 0.7620] & [ -0.7571, 0.7682] \\
			[0.0703 ,  0.4736]  & [-0.4739,0.7620 ] & [ 1.5868, 3.9349] & [ -1.3594, 1.0586] \\
			[-0.2578,  0.1985] & [-0.7571, 0.7682 ] & [-1.3594, 1.0586] & [  2.4684, 9.1016]
		\end{pmatrix}
	\end{equation}
	
	\noindent
}


	\begin{figure}[h]
		\centerline{
				\subfigure[MAE versus $N$. \label{ME_CAGO_EN_TODOS_A}]
{\includegraphics[width=8cm]{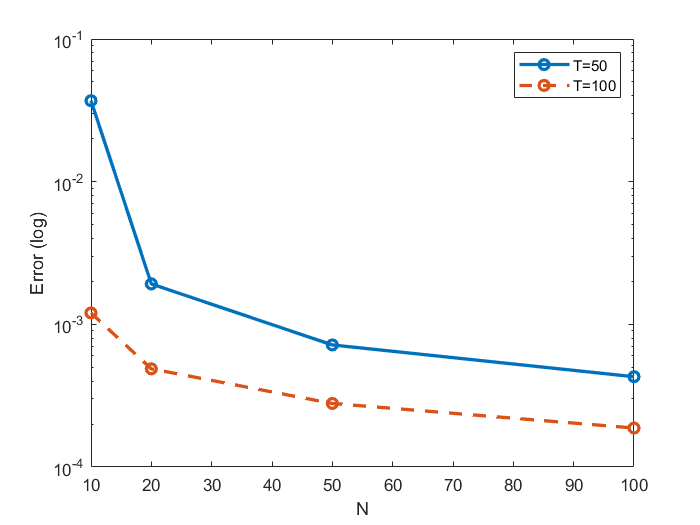}}
		\subfigure[{MAE versus $T$.}  \label{ME_CAGO_EN_TODOS_B}]{\includegraphics[width=8.2cm]{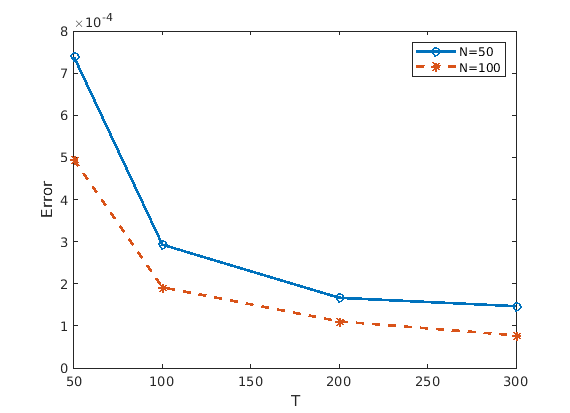}}}
		\caption{{Multi-output example.} MAE in the estimation of the MAP of the posterior (considering ${\bm \theta}_{\texttt{true}}$ as the groundtruth) by ATAIS; {\bf (a)} versus the number of particles $N$   { with fixed $T$ and {\bf (b)}  versus iterations $T$ with fixed $N$.}}\label{NO_CITADA_ME_CAGO_EN_TODOS_LOS_M_DE_ERNESTO}
	\end{figure}
	
	\begin{figure}[H]
\centerline{
\subfigure[{Keeping constant the product $NT$. As $T$ grows, $N$ decreases, e.g., in $T=500$, we have $N=40$, and in $T=2000$ we have $N=10$.}\label{Tu_puta_madre}]{ \includegraphics[width=7.1cm]{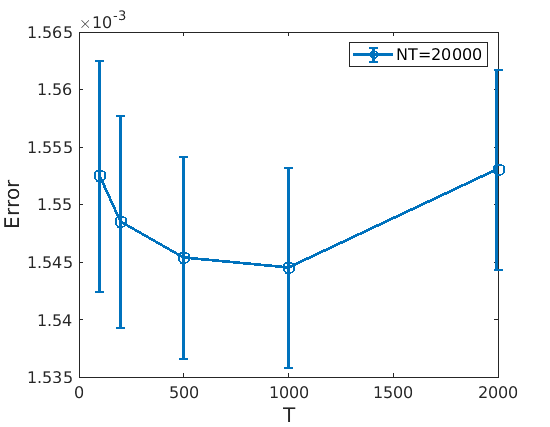}}
	\subfigure[$\widehat{\btheta}_{\texttt{MAP}}$ with different schemes.\label{Tu_puta_madre2}]{ \includegraphics[width=7.3cm]{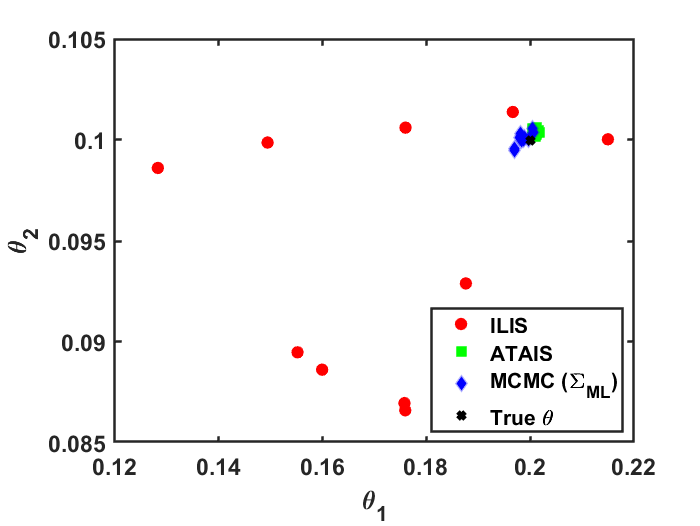}}	
}
	\caption{{Multioutput example. {\bf (a)} Complete MAE in  estimating the $\btheta_{\texttt{MAP}}$ and the $\bSigma_{\texttt{ML}}$, keeping fixed the number of posterior evaluations $NT=20000$, with $T\in \{100, 200, 500, 1000, 2000\}$, hence $N\in\{200, 100, 40, 20, 10\}$.} {\bf (b)} Estimations of $\btheta_{\texttt{MAP}}$. The green squares represent the estimation with ATAIS; the red circles depict the estimations of ILIS with a covariance matrix generated from a Wishart distribution. The blue diamonds represent the estimations of the single MH addressing a conditional posterior as invariant density, where ${\bm \Sigma}$ is fixed to the value of estimation $\bm{\Sigma}_{\texttt{ML}}$ obtained by ATAIS. The black cross depicts ${\bm \theta}_{\texttt{true}}$.}
	\label{fig:atais-ilis_2} 
\end{figure} 
	

		\begin{table}[!h]
			\centering
			\caption{{Multi-output example: MAE averaged over 1000 simulations of ATAIS for estimating the $\bm{\theta}_{\texttt{true}}$, $\bSigma_\texttt{ML}$ (its groundtruth is computed as in Eq. \eqref{MatrixTRUEeq2}), and the complete MAE in the whole space. }}
			\label{tab:mae-multioutput-T-N}
			\vspace{0.2cm}
			
			\begin{tabular}{l|lll|l|lll|}
				\cline{2-4} \cline{6-8}
				& \multicolumn{3}{l|}{{\bf Varying in $N$ (with fixed $T=50$)}}                           &    & \multicolumn{3}{l|}{{\bf Varying in $T$ (with fixed $N=100$)}}                          \\ \hline
				\multicolumn{1}{|l|}{$N$}   & \multicolumn{1}{l|}{$\btheta_{\texttt{true}}$} & \multicolumn{1}{l|}{$\bSigma_{\texttt{ML}}$} & {\bf Complete MAE}  & $T$  & \multicolumn{1}{l|}{$\btheta_{\texttt{true}}$} & \multicolumn{1}{l|}{$\bSigma_{\texttt{ML}}$} & {\bf Complete MAE}   \\ \hline
				\multicolumn{1}{|l|}{5}   & \multicolumn{1}{l|}{0.2325} & \multicolumn{1}{l|}{0.7666} & 0.7073 & 5  & \multicolumn{1}{l|}{0.2100} & \multicolumn{1}{l|}{0.4648} & 0.4365 \\ \hline
				\multicolumn{1}{|l|}{12}  & \multicolumn{1}{l|}{0.0102} & \multicolumn{1}{l|}{0.0136} & 0.0132 & 10 & \multicolumn{1}{l|}{0.1219} & \multicolumn{1}{l|}{0.1293} & 0.1285 \\ \hline
				\multicolumn{1}{|l|}{25}  & \multicolumn{1}{l|}{0.0013} & \multicolumn{1}{l|}{0.0026} & 0.0024 & 20 & \multicolumn{1}{l|}{0.0355} & \multicolumn{1}{l|}{0.2663} & 0.2407 \\ \hline
				\multicolumn{1}{|l|}{50}  & \multicolumn{1}{l|}{0.0012} & \multicolumn{1}{l|}{0.0023} & 0.0021 & 30 & \multicolumn{1}{l|}{0.0015} & \multicolumn{1}{l|}{0.0031} & 0.0029 \\ \hline
				\multicolumn{1}{|l|}{100} & \multicolumn{1}{l|}{0.009}  & \multicolumn{1}{l|}{0.0017} & 0.0025 & 50 & \multicolumn{1}{l|}{0.0010} & \multicolumn{1}{l|}{0.0021} & 0.0019 \\ \hline
			\end{tabular}
		\end{table}

\newpage
\clearpage

\begin{figure}[!htp]
		
		\centering
		\includegraphics[width=0.5\textwidth]{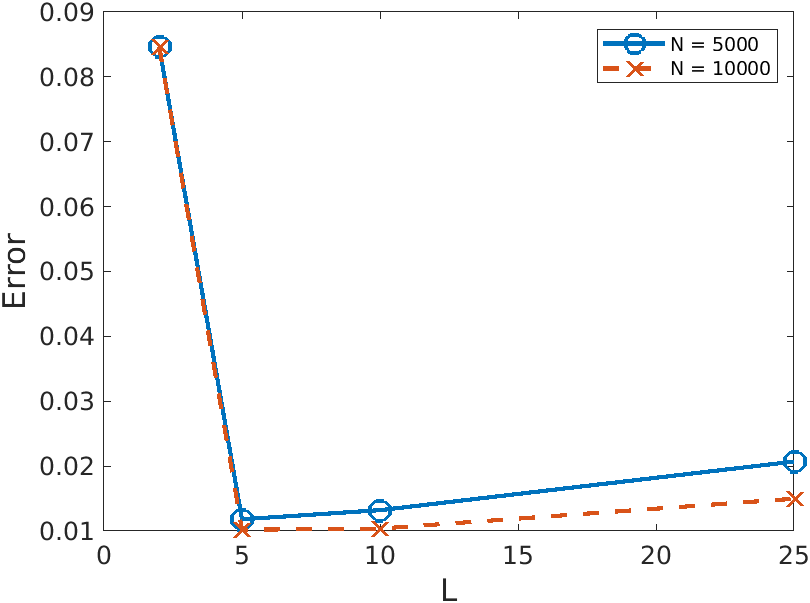}
		
		\caption{Testing the use of mini-batches in the multi-output model in Section \ref{SectMO}: MAE in estimating the full parameter space $\{{\bm \theta},{\bm \Sigma}\}$ with $R=50$ (number of data), $L\in \{2, 5, 10, 25\}$ (number of data in the batches), $T=\frac{R}{L}\in \{50, 10, 5, 2\}$ (number of iterations which coincides with the number of batches). Note that the curve of  $N=5000$ is depicted with a solid line, and the results with $N=10^4$ are shown with a dashed line. }
		\label{fig:batches}
	\end{figure}

{
\subsection{Use of mini-batches}\label{MiniBSEct}

	In this experiment, we test the possible use of mini-batches in ATAIS. 
	More specifically, we test the second strategy in Section \ref{BatchSect}.   We consider the multi-output model in Section \ref{SectMO}. We consider different number of data  in the batches, $L\in \{2, 5, 10, 25\}$.  We have $R=50$ observations.
 Following Section \ref{BatchSect}, the data are randomly divided into $\frac{R}{L}$ disjoint subsets, with $L$ elements in each batch. Namely, we have exactly $\frac{R}{L}$ batches. The number of iterations in ATAIS, $T$, is also equal to $\frac{R}{L}$, which forces to take fewer ATAIS steps as we increase the size of the subsets. Hence, we have $T=\frac{R}{L}$ that represents both the number of iterations and the number of batches.
 Since in this example the total number of observations is $R=50$, we have $T \in\{25, 10, 5, 2\}$ (according to the values of $L$ given above). The number of particles in each iteration of ATAIS is set to $N\in\{5000,10^4\}$.
 \newline
Figure \ref{fig:batches} shows the MAE in estimating the full parameter space $\{{\bm \theta},{\bm \Sigma}\}$ (considering $\bSigma_{\texttt{ML}}$ in Eq. \eqref{MatrixTRUEeq2} as groundtruth), averaged over $10^5$ independent runs.  We can see that the MAE is always very small in all cases.
Hence, ATAIS works well even using mini-batches. Moreover, since the use of mini-batches has a tempering effect that fosters the exploration of the space, ATAIS works even better with batches when the number of data in each batch, $L$, is not too small (in Figure \ref{fig:batches}, the choice $L=2$ degrades slightly  the performance).
	

	%
}
	
 {	
\subsection{Testing ATAIS in models with a Student's-t noise}\label{StudentSect}

As described in Section \ref{OtherNoiseSect}, ATAIS  can be also applied when the noise perturbation in the observation model follows an elliptical distribution. In this  experiment, we consider a noise distribution with heavier tails with respect to the Gaussian case. More specifically, we consider a noise variable $\textbf{v}_r$ in Eqs. \eqref{EqObsModel_all}-\eqref{NoiseVariable} is distributed as a multivariate Student's-t distribution with 10 degrees of freedom. In this case, the estimation of ${\bm \Sigma}$ is obtained by the iterative process described in Section \ref{OtherNoiseSect}.
 We test the case of the Student's-$t$ noise in both models described in the experiments above: the localization model in Section \ref{LocSect} and the multi-output model in Section \ref{SectMO}.
 \newline
  In Table \ref{tab:N-cambia-T-fijo-Student}, we provide the averaged MAE over $10^3$ runs in the estimations of ATAIS for $\btheta_{\texttt{true}}$, and $\bSigma_{\texttt{ML}}$, computed as in Eq. \eqref{MatrixTRUEeq2}. We set the number of iterations to $T=50$ and the number of particles $N$ is taken as $\{5, 12, 25, 50, 100\}$. Table \ref{tab:N-cambia-T-fijo-Student} shows the results for the localization model in Section \ref{LocSect} and the multi-output model in Section \ref{SectMO} both with  the Student's-$t$ noise.   Again we can observe that ATAIS provides good results. The MAE decreases as the number of particles $N$ increases, as expected. It is worth noting that in this scenario with a noise with a heavier tail the estimation of the matrix $\bSigma_{\texttt{ML}}$ seems to be more difficult than with the Gaussian noise scenario. 
  
}
\begin{table}[H]
	
	\centering
	\caption{Experiment with Student's-t noise: MAE averaged over 1000 simulations of ATAIS for estimating the $\bm{\theta}_{\texttt{true}}$ and $\bm{\Sigma}_\texttt{ML}$ (its groundtruth is computed as in Eq. \eqref{MatrixTRUEeq2}). The number of iterations is fixed to $T=50$.}
	\label{tab:N-cambia-T-fijo-Student}
	\vspace{0.2cm}
	
	\begin{tabular}{l|lll|lll|}
		\cline{2-7}
		                          &                           \multicolumn{3}{c|}{{\bf  Localization model with Student's-t noise}}                            &                           \multicolumn{3}{c|}{{\bf Multi-output model  with Student's-t noise}}                            \\ \hline
		\multicolumn{1}{|l|}{$N$} & \multicolumn{1}{l|}{$\bm{\theta}_{\texttt{true}}$} & \multicolumn{1}{l|}{$\bm{\Sigma}_{\texttt{ML}}$} & {\bf Complete MAE} & \multicolumn{1}{l|}{$\bm{\theta}_{\texttt{true}}$} & \multicolumn{1}{l|}{$\bm{\Sigma}_{\texttt{ML}}$} & {\bf Complete MAE} \\ \hline
		\multicolumn{1}{|l|}{5}   & \multicolumn{1}{l|}{0.1732}                        & \multicolumn{1}{l|}{12.3840}                     & 10.1639            & \multicolumn{1}{l|}{0.1481}                        & \multicolumn{1}{l|}{0.3517}                      & 0.3291             \\ \hline
		\multicolumn{1}{|l|}{12}  & \multicolumn{1}{l|}{0.0752}                        & \multicolumn{1}{l|}{2.4801}                      & 2.0428             & \multicolumn{1}{l|}{0.0747}                        & \multicolumn{1}{l|}{0.1157}                      & 0.1111             \\ \hline
		\multicolumn{1}{|l|}{25}  & \multicolumn{1}{l|}{0.0470}                        & \multicolumn{1}{l|}{0.6286}                      & 0.5229             & \multicolumn{1}{l|}{0.0484}                        & \multicolumn{1}{l|}{0.0579}                      & 0.0568             \\ \hline
		\multicolumn{1}{|l|}{50}  & \multicolumn{1}{l|}{0.0372}                        & \multicolumn{1}{l|}{0.1094}                      & 0.0963             & \multicolumn{1}{l|}{0.0337}                        & \multicolumn{1}{l|}{0.0377}                      & 0.0366             \\ \hline
		\multicolumn{1}{|l|}{100} & \multicolumn{1}{l|}{0.0327}                        & \multicolumn{1}{l|}{0.0753}                      & 0.0676             & \multicolumn{1}{l|}{0.0246}                        & \multicolumn{1}{l|}{0.0284}                      & 0.0279             \\ \hline
	\end{tabular}
\end{table}


\subsection{Application to a biology system {- dimension of the space $D=7$}}\label{BioExample}
\begin{figure}[!h]
	\subfigure[{ $\widehat{{\bm \theta}}_{\texttt{MAP}}^{(t)}$}]{\includegraphics[width=0.47\textwidth]{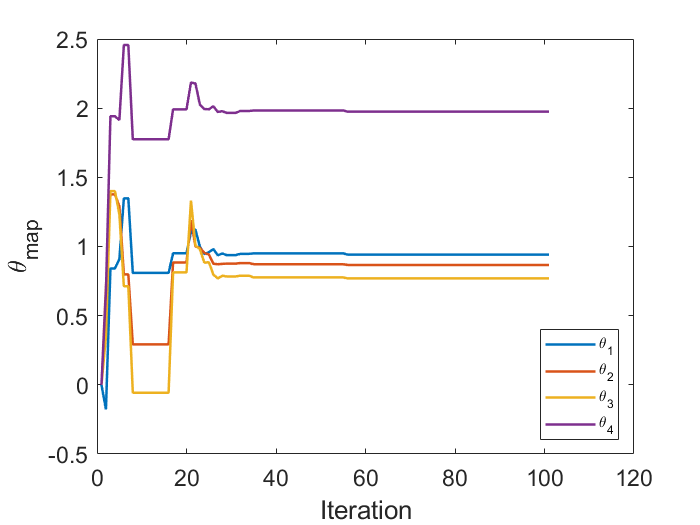}}
	\subfigure[{$\widehat{{\bm \Sigma}}_{\texttt{ML}}^{(t)}$}]{\includegraphics[width=0.47\textwidth]{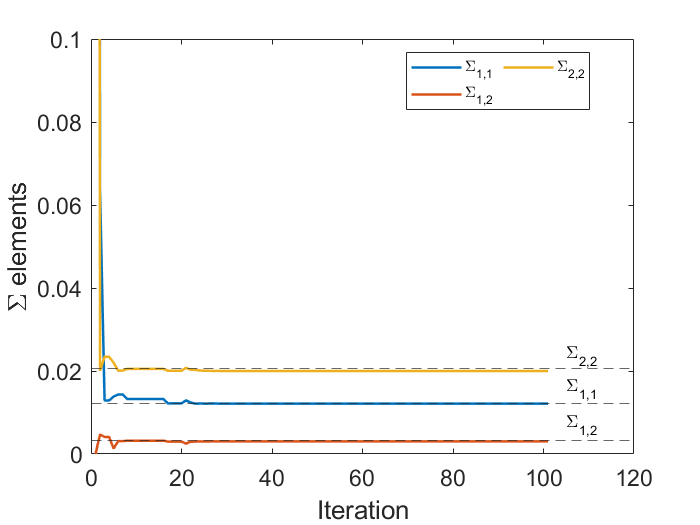}}
	{\caption{Application to a system biology example. Evolution of the components in the estimators $\widehat{{\bm \theta}}_{\texttt{MAP}}^{(t)}$ and $\widehat{{\bm \Sigma}}_{\texttt{ML}}^{(t)}$ versus $t$, in one run of ATAIS algorithm.}}\label{MECAGO_OTRA_VEZ}
\end{figure}

In this third  example of application, we focus on an inference problem in a  biology system \cite{BayesianParameterEstimationForDynamicalModelsInSystemsBiology}. We aim to make inference on the covariance matrix of the observations of a model. This model represents a physiological system with two states and one input variable. The model is governed by a set of four parameters denoted as $\bm{\theta}=[k_{12}, k_{21}, k_{1e}, b]^{\top}$. The dynamics of the system is ruled by the following differential equations. using the notation in Eq. \eqref{NotationAqui}:
\begin{equation}\label{equ:ode_sys}
	\begin{split}
		\dfrac{df_1}{d\tau}  = & -(k_{1e} + k_{12})\cdot f_1(\bm{\theta},\tau) + k_{21}\cdot f_2(\bm{\theta},\tau) + b\cdot u(\tau),\\
		\dfrac{df_2}{d\tau}  = & k_{12} \cdot f_1(\bm{\theta},\tau)  - k_{21}\cdot f_2(\bm{\theta},\tau),
	\end{split}
\end{equation}
where
\begin{equation}\label{key}
	u(\tau) = \begin{cases}
		\tau + 0.5 & \text{if } 0\leq \tau \leq 1\\
		1.5 e^{1-\tau}  & \text{if } \tau>1,
	\end{cases}
\end{equation}
is an input of the system. We set the true parameters to $\bm{\theta}_{\texttt{true}}=[1,1,1,2]^{\top}$ and
$$\bm{\Sigma}_{\texttt{true}}=
\begin{pmatrix}
	1 & 0.9 \\
	0.9 & 2
\end{pmatrix}.$$
The prior for the components of $\btheta$ was set as uniform in the interval $[0, 5]$ (for all the components) as suggested in \cite{BayesianParameterEstimationForDynamicalModelsInSystemsBiology}.
The system of differential equations above has no analytically close solution. Hence,
	the generation of the data according to the system  is obtained   using a Runge-Kutta Matlab solver \cite{Moler2004}.  We assume that the solution is perturbed by Gaussian noise according to the  distribution $\mathcal{N}(0, \bm{\Sigma}_{\texttt{true}})$. Then, 100 data points are sampled at equidistant times from the solution of the system \eqref{equ:ode_sys}.
	\newline
\newline	
{\bf Remark.}	Note that, in this experiment, the function $\textbf{f}$ is evaluated only approximately since it is the solution of the system \eqref{equ:ode_sys} which cannot be evaluated exactly. Thus, even considering the vector of true values $\bm{\theta}_{\texttt{true}}$,  we need to approximate $\textbf{f}(\bm{\theta}_{\texttt{true}},\tau)$ by a differential equation solver (discretizing it) obtaining $\widehat{\textbf{f}}(\bm{\theta}_{\texttt{true}},\tau)$. 
\newline
\newline	
	In order to make inference we use ATAIS with $ N=300 $ particles  during $ T=100 $ iterations. The initial value for the mean of the proposal is set at $[0, 0, 0 ,0]$ and for the covariance matrix is $6\textbf{I}_4$. The initial estimate for the covariance matrix of the data, $\widehat{\bSigma}_{\texttt{ML}}^{(0)}$, was the identity, $\textbf{I}_2$.
{The convergence in one run of ATAIS estimators can be observed in Figure \ref{MECAGO_OTRA_VEZ}. }

\noindent	
In the second part of ATAIS, we generate $J=1000$ matrices from a Wishart distribution (degrees of freedom, $\nu$ = 100) with reference matrix equal to $\widehat{\bSigma}_{\texttt{ML}}^{(T)}$, the final estimate of the first part of ATAIS. To each of the generated matrices we assign a weight following Eq \eqref{equ:sigma weights}. After performing resampling. Applying resampling according to the weights $\{\bar{\lambda}_j\}$ we calculate the percentiles 0.025 and 0.975 to get a 95\% credible interval (averaged over 50 independent runs) in Eq \eqref{equ:interval edo}. It can be seen that our interval contains the true components of the covariance matrix.
	\begin{equation}\label{equ:interval edo}
	\begin{pmatrix}
		[0.7820, 1.3876] & [0.7099, 1.4614] \\
		[0.7099, 1.4614] & [1.6518, 2.8760]
	\end{pmatrix}	
	\end{equation}

\noindent
As a final remark, in this example, the posterior has a particularly flat shape in some regions, so that many values of the parameters $\btheta$  provide acceptable results for the solution of the system \eqref{equ:ode_sys}, even if the distance to the $\btheta_{\texttt{true}}$ could be large.


{
\subsection{Application to graph topology estimation - dimension of the space $D=59$ }\label{GraphExamples}
In this section, the goal is to apply ATAIS to estimate a graph topology with $10$ nodes.  The graph is  represented by a $10\times 10$ graph shift operator that is usually considered a sparse matrix \cite{MarquesStationaryGraph,uhler2018gaussian}. In this experiment, we consider that the precision matrix $\mathbf{P} = \bSigma^{-1}$ is the graph shift operator. Thus, we assume $\mathbf{P}$ to be a sparse matrix and hence represent the graph \cite{MarquesStationaryGraph,uhler2018gaussian}.   If the nodes $i$ and $j$ are connected, then it follows that $\mathbf{P}_{i,j} \neq 0$.  
  Estimating the graph topology consists mainly of knowing the connections between the nodes, which is determined by the matrix $\mathbf{P} = \bSigma^{-1}$. The goal is to apply ATAIS to estimate the graph topology. The corresponding adjacent matrix  (binary matrix showing the connection, that is the groundtruth in this experiment) is depicted in Figure \ref{subfig:graph-true-connection}. Furthermore, we assume the following multi-output nonlinear model, 
\begin{equation}\label{equ:graph f}
	\begin{array}{rcl}
		f_{1}(\btheta) &=& -\theta_4 \tau + 5 \theta_1^2 \\
		f_{2}(\btheta) &=& 2\theta_3 \sin-\theta_2 \tau \\
		f_{3}(\btheta) &=& \theta_1 - \theta_3 + \theta_1\cos(2 \tau) \\
		f_{4}(\btheta) &=& 3\theta_4 + 3\theta_2 + \theta_1 \exp(0.1 \tau) \\
		f_{5}(\btheta) &=& \theta_3^2 - 2 \theta_1 + 3\theta_2 - \exp(0.8 \theta_3) \exp(1 - \tau) \\
		f_{6}(\btheta) &=& 5 (\theta_4 + \theta_3) - \theta_2 \log(1 + 2 \tau) \\
		f_{7}(\btheta) &=& 3 \theta_2 - 0.2 \tau \sin(\theta_3) \\
		f_{8}(\btheta) &=& 3\theta_1 + 5 \theta_3 - 20\sin(\theta_4) \cos(2 \tau +\dfrac{\pi}{4}) \\
		f_{9}(\btheta) &=& \theta_2 + 4\theta_4 + 5 \exp\left (\dfrac{1}{1 + \theta_3}\right ) \tau \\
		f_{10}(\btheta) &=& 5\theta_1 + 10\theta_3 - 5\theta_4 \sin(\tau),
	\end{array}
\end{equation}
where  $f_{r,i}(\btheta)$ is $i$-th signal contaminated by Gaussian noise and is emitted by at the $i$-th node. The model in Eq. \eqref{equ:graph f} is characterized by a  vector parameter $\btheta \in \mathbb{R}^4$ with the following true values  $\btheta_{\texttt{true}}=[0.5, 2, 5, 3]^{\top}$.  We assume an improper uniform prior over  $\btheta$.
Note that, in this scenario, we have $M=4$ and $K=10$, hence  the complete space has dimension,
$$
D=M+\frac{K(K+1)}{2}=59.
$$
We estimate $\bSigma$ with ATAIS, then compute $\mathbf{P} = \bSigma^{-1}$ and apply a threshold value of $0.3$  for obtaining a corresponding binary matrix (entries of  $\mathbf{P}$ smaller than $0.3$ are set to 0, whereas entries of  $\mathbf{P}$ greater than $0.3$ are set to 1). More sophisticated strategies could be considered, for instance, using ATAIS jointly with a graphical lasso approach. We set $T=10$  and $N=20000$ and consider $500$ independent runs.
For the adaptation of the covariance of the proposal density in Eqs. \eqref{SigmaEqNoTable}-\eqref{SigmaEqTable}, we apply a cyclic/periodic strategy in Eq. \eqref{DeltStrategy}
 \begin{gather}
\delta_{t+1}=  \left\{
  \begin{split}
&a \ \delta_{t}, \qquad   \mbox{ if }\delta_t \geq  \delta_{\texttt{min}}, \\
& \delta_{0},    \mbox{ }\mbox{ }\mbox{ }\qquad \mbox{ if }\delta_t < \delta_{\texttt{min}},
 \end{split}
 \right.
  \end{gather}
with $\delta_{0}=1$, $a=0.1$ and $\delta_{\texttt{min}}=0.05$.  
\newline
The results in estimation of $\mathbf{P} = \bSigma^{-1}$ are given in percentages in Figure \ref{GrafoFiguras}. 
 In 69.7$\%$ of the runs, we obtain exactly the groundtruth matrix in {\bf (a)}. In the rest of the runs, we get very similar adjacency matrices, shown in {\bf (b)}-{\bf (c)}.
 \newline
{Figure \ref{otraLucaSampleImp} depicts the evaluations of the final conditional log-posterior of the samples $\btheta_t^{(n)}$  such that $\bar{w}_t^{(n)}\geq 1/N$ at each iteration. Hence, the number of these relevant samples varies with iteration $t$ and with the specific run, depending also on the adaptation of the proposal and its covariance matrix  in Eqs. \eqref{SigmaEqNoTable}-\eqref{SigmaEqTable}. For instance, in Figure \ref{otraLucaSampleImp1}, we can observe the evaluation in 3 different runs where the number of extremely relevant samples - with weights such that $\bar{w}_t^{(n)}\geq \frac{1}{N}$ - is $70$, $73$ and $111$, respectively. These numbers (of the extremely relevant samples) can be considered small, but actually they are very well-located as shown in  Figure \ref{otraLucaSampleImp2}, which illustrates the histograms of the four components of these relevant samples. Note that the histograms are localized around the true values of $\btheta_{\texttt{true}}=[0.5, 2, 5, 3]^{\top}$. Indeed, the values of the log-posterior evaluations are close  to the log-posterior evaluation at $\btheta_{\texttt{true}}$.
 Computing an averaged (standard) ESS measure $\widehat{ESS}=\frac{1}{T} \sum_{t=1}^T \mbox{ESS}(t)$ (averaged over the iterations, and also over the different runs) where $\mbox{ESS}(t)=1/\sum_{n=1}^N \bar{w}_t^{(n)}$   \cite{ESSarxiv16}, we obtain $\frac{\widehat{ESS}}{N}\approx 0.10$, i.e., $10 \%$ of the $N$ particles. However, we can see they are well located, in fact, the MAE in estimation of ${\bm \theta}$ is only 0.01. Therefore, ATAIS provides very good performance, remarking that the dimension of the space is $D=59$.   }



\begin{figure}[htp]
{
	\centering
	\subfigure[True matrix, obtained in estimation 69.7$\%$ of times. \label{subfig:graph-true-connection}]{\includegraphics[width=5cm]{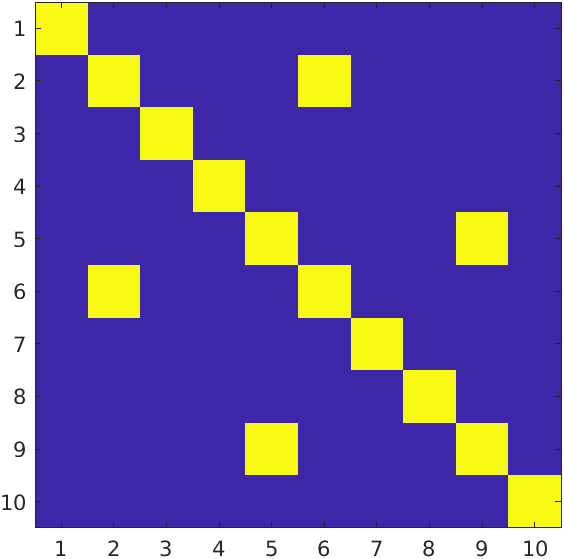}}
	\subfigure[Obtained in estimation 22.1$\%$ of times. \label{subfig:graph-estimated-connection}]{\includegraphics[width=5cm]{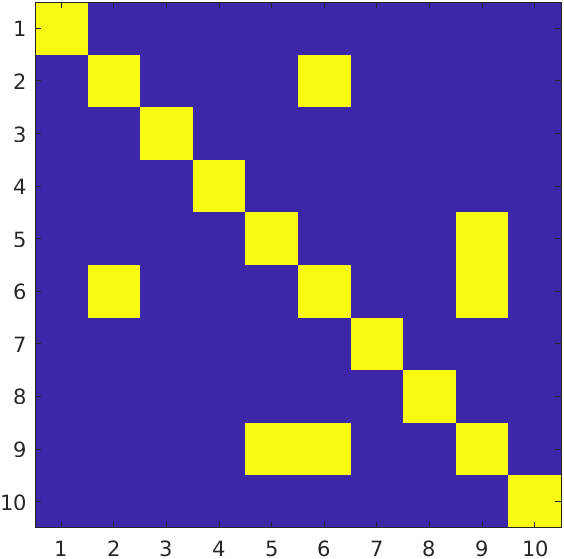}}
	\subfigure[Obtained in estimation 8.2$\%$ of times. \label{subfig:graph-other-estimated-connection}]{\includegraphics[width=5cm]{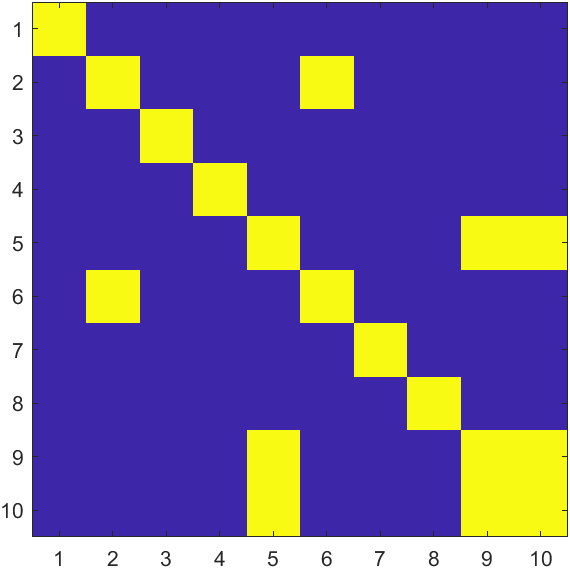}}
	\caption{Graph topology estimation with ATAIS algorithm (with $T=10$ and $N=2\cdot 10^4$). In the 69.7$\%$ of the runs, we obtain exactly the groundtruth matrix in {\bf (a)}. In the rest of other runs, we get very similar adjacency matrices, shown in {\bf (b)}-{\bf (c)}.}
	\label{GrafoFiguras}}
\end{figure}

\begin{figure}[htp]
\subfigure[\label{otraLucaSampleImp1}]{\includegraphics[width=8cm]{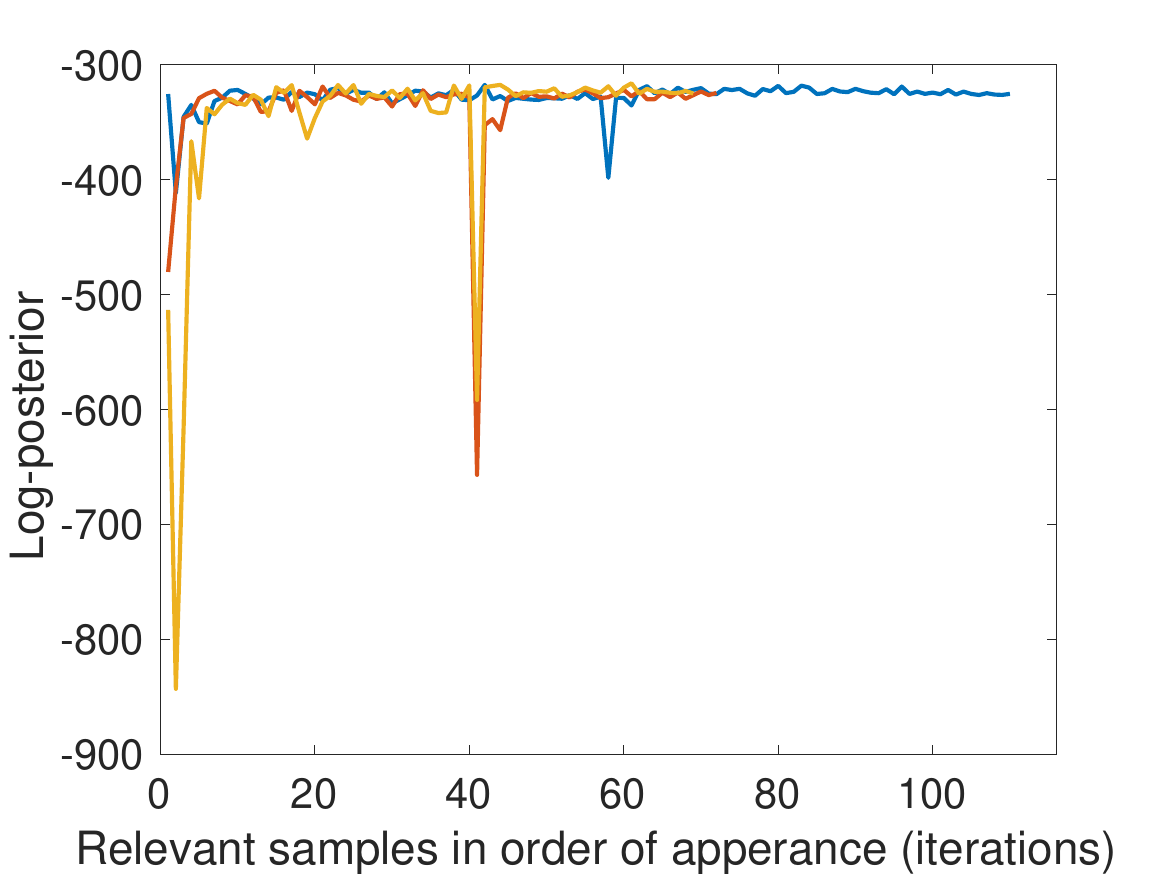}}
\subfigure[\label{otraLucaSampleImp2}]{\includegraphics[width=8cm]{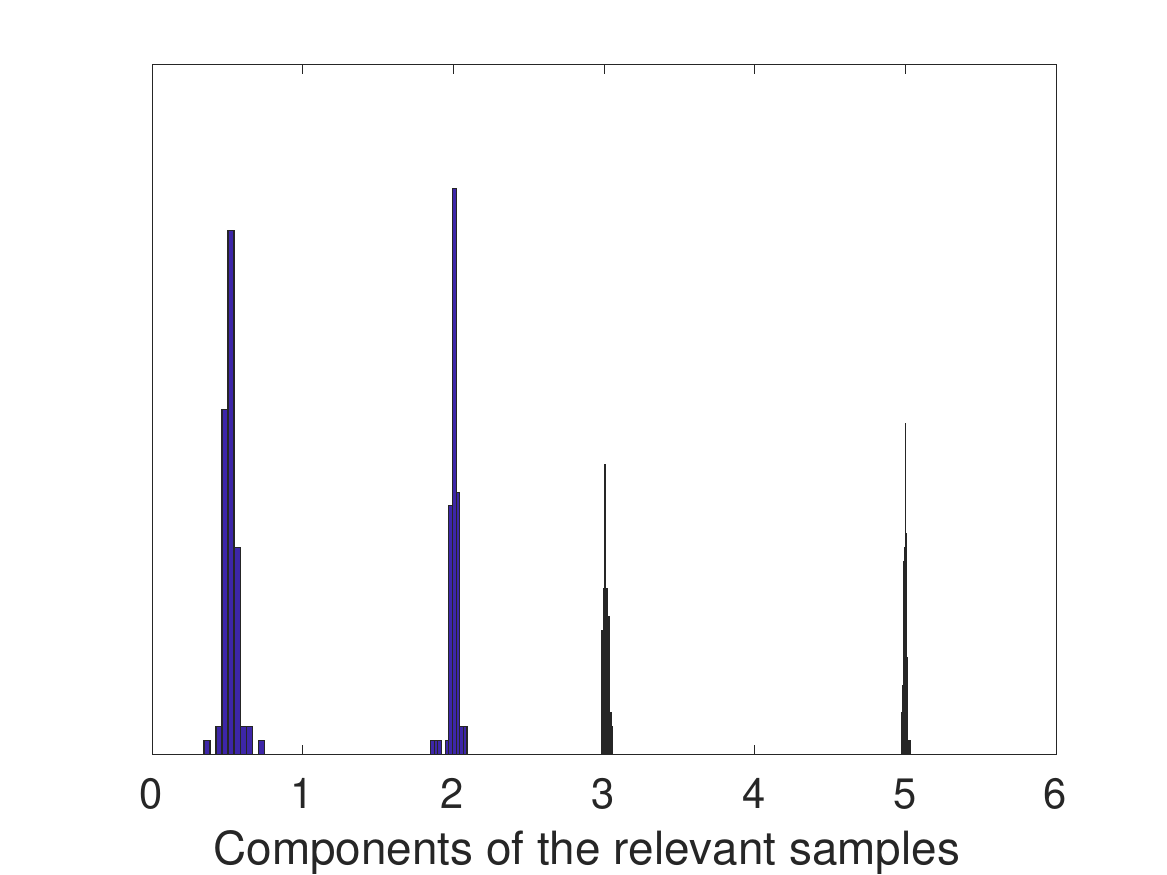}}
	\caption{{ {\bf (a)} Evaluations of the final conditional log-posterior of the samples $\btheta_t^{(n)}$  such that $\bar{w}_t^{(n)}\geq 1/N$ at each iteration. {\bf (b)} Histograms of the four components of the relevant samples. They are localized around the true values of $\btheta_{\texttt{true}}=[0.5, 2, 5, 3]^{\top}$.}}\label{otraLucaSampleImp}
\end{figure}

}

\newpage
{ \subsection{Comparison with other benchmark schemes}\label{ComparisonSect}

First of all, it is important to remark that, in all the numerical experiments above, the application of other Monte Carlo approaches directly on the joint space $\{{\bm \theta},{\bm \Sigma}\}$ often produces much higher errors in estimation, which makes it difficult to compare and visualize with respect to the results obtained by ATAIS. 
In this section, we make an effort to compare with other techniques.
Specifically, we apply  different MCMC algorithms on the complete parameter space $\btheta, \bSigma$, in order to compare to ATAIS.  Since ATAIS can be applied even when ${\bf f}$ is not differentiable, in other to perform a fair comparison, we compare with benchmark algorithms in the literature that do not use the gradient information. We consider the multi-output model with Gaussian noise in Section \ref{SectMO}. Hence, as shown in Section \ref{SectMO}, the dimension of the complete space is $D=12$. 
\newline
\newline
Firstly, we apply a Metropolis-Hastings (MH) algorithm, testing different parameters. We use different random walk Gaussian proposal densities for the $\btheta$ with covariance matrix $a \mathbf{I}_M$ with $a\in \{0.1, 1,10\}$  and for the proposal for generating $\bSigma$ is a random walk Wishart with $\nu\in \{5,50\}$ degrees of freedom, with a length of the chain of $T=2\cdot 10^4$. Secondly, we apply an adaptive version of the previously described MH schemes where the random walk Gaussian proposal density with respect to $\btheta$ is tuned using the empirical mean and variance of the generated past samples \cite{Robert04,LuengoSurvey}. Then, the parameter $a$ is auto-tuned.
Finally, we also applied also MH-within-Gibbs sampler (that is a more adequate scheme for sampling tight posteriors in high-dimensional spaces) where, for each internal Gibbs iteration for sampling from each full-conditional, we apply $10$ MH internal steps  \cite{StickyGibbs}. The internal MH considers random walk Gaussian proposal pdf with unit variance. Moreover, since the elements of the matrix $\bSigma$  required to fulfill certain constraints we discard the matrices that are not positive definite. Hence, this scheme is much more costly than ATAIS and the other MCMC algorithms above.
 At each run, the initializations of the chains of all the MCMC schemes (as in ATAIS) described above are chosen randomly around a vector of zeros for ${\bm \theta}$ and an identity matrix for ${\bm \Sigma}$ (with a standard deviation of $4$; for ${\bm \Sigma}$, we always consider an initial diagonal matrix with the same random element in the diagonal). 
 \newline
  Table  \ref{tab:rw-complete-multioutput} shows the MAE (averaged over $10^3$ independent runs) for different schemes in the multi-output example. We include also the MAE corresponding to ATAIS for facilitating the comparison.  Note that, we consider always more evaluations of the model for the MCMC schemes than ATAIS, except for the last line of ATAIS. In any case, ATAIS clearly outperforms the rest of the compared techniques.  
  
%

\begin{table}[!h]
	
	\centering
	\caption{Comparison with other techniques in the multi-output model of Section \ref{SectMO}: MAE in the complete space by different methods.  }
	\label{tab:rw-complete-multioutput}
	\begin{tabular}{|c|c|c||c|}
		\hline
		           {\bf Method}            &       {\bf Parameters}        &        {\bf Posterior evaluations}        & {\bf MAE}   \\ \hline\hline
		                MH                 & $a=0.1, \nu=5, T=2\cdot10^4$  & \multirow{6}{*}{$T=2\cdot10^4$} &  154.71     \\
		                MH                 & $a=0.1, \nu=50, T=2\cdot10^4$ &                                 &  0.6635     \\
		                MH                 &  $a=1, \nu=5, T=2\cdot10^4$   &                                 &  6.0458     \\
		                MH                 &  $a=1, \nu=50, T=2\cdot10^4$  &                                 &  0.8170    \\
		                MH                 &  $a=10, \nu=5, T=2\cdot10^4$  &                                 &  3.6669     \\
		                MH                 &  $a=10, \nu=50, T=2\cdot10^4$  &                                 &  1.4905     \\ \hline\hline
		           adaptive MH             &            $\nu=5$     $T=2\cdot10^4$       &      \multirow{2}{*}{$T=2\cdot10^4$}                            &       2.9072         \\
		           adaptive MH             &            $\nu=50$    $T=2\cdot10^4$        &                       &     0.7481         \\ \hline\hline
		             MH-within-Gibbs                &     $10$ internal MH steps, $T=2\cdot10^4$         &   $T=2\cdot10^4$                                &          0.1351    \\ \hline\hline
		              ATAIS                &        $N=100$, $T=5$         &            $NT=500$             &  0.4365     \\
		              ATAIS                &        $N=100$, $T=20$        &            $NT=2000$            &  0.2407     \\
		              ATAIS                &       $N=100$, $T=100$        &            $NT=10^4$            &  0.0015     \\ 
		              ATAIS                &  $N=100, T=200$               &            $ NT=2\cdot10^4$     &  0.0012    \\ \hline
	\end{tabular}
\end{table}

}
%

\section{Conclusions}\label{conclSect} 

In this work, we have introduced an adaptive importance sampling (AIS) method for robust inference in complex Bayesian inversion problems with unknown parameters ${\bm \theta}$ of the non-linear mapping and unknown covariance matrix ${\bm \Sigma}$ of the noise perturbation. The variables of interest are split in two blocks, the parameters  ${\bm \theta}$ of the non-linear model and the covariance matrix ${\bm \Sigma}$, are handled in different ways. The main proposed inference scheme is divided into two main parts. The first part is devoted to approximating a conditional posterior ${\bm \theta}$  given the data and the maximum likelihood estimator of  ${\bm \Sigma}$.  This first part allows of finding regions of high probability about ${\bm \theta}$ and ${\bm \Sigma}$ (working alternately in subsets of the complete space, with reduced dimensions).
\newline
In the second part, a Bayesian approach is also performed over ${\bm \Sigma}$ re-using and  re-weighting the samples of ${\bm \theta}$ previously generated.  Then, an approximation of the complete posterior of $\{{\bm \theta},{\bm \Sigma}\}$ is provided. This second part does not require of additional evaluation of the possibly costly non-linear vectorial model ${\bf f}$. 
 The resulting scheme is a robust inference approach for Bayesian inversion, based on an adaptive importance sampler that addresses a sequence of different conditional posteriors and a post-process that allows a Bayesian inference also over $ {\bm\Sigma}$. { Several variants and computational details have been discussed.} Additionally, a simpler compelling scheme, called ILIS, has been introduced in order to compare with ATAIS.
 { Several numerical experiments have been presented.} We can observe the good performance obtained by ATAIS,  providing a complete Bayesian analysis of the complete space $\{{\bm \theta},{\bm \Sigma}\}$, { and outperforming other benchmark algorithms}.




%

\section*{Acknowledgments}
The work was partially supported by the Young Researchers R\&D Project, ref. num. F861 (AUTO-BA-GRAPH) funded by the Community of Madrid and Rey Juan Carlos University, and by Agencia Estatal de Investigaci{\'o}n AEI (project SP-GRAPH, ref. num. PID2019-105032GB-I00).

\begin{table}[h!]
\caption{ATAIS: an adaptive IS scheme with a sequence of adaptive target pdfs  \label{AIS_AutoTemp}}
\begin{tabular}{|p{0.95\columnwidth}|}
    \hline
\begin{enumerate}
\item {\bf Initializations:} Choose $N$, ${\bm \mu}_1$, ${\bm \Lambda}_1$, $\widehat{{\bm \Sigma}}_{\texttt{ML}}^{(0)}$, and set $\pi_{\texttt{MAP}}=0$. Recall $\pi_t({\bm \theta}) \propto p({\bm \theta}|{\bf Y},\widehat{{\bm \Sigma}}_{\texttt{ML}}^{(t-1)})$.
 \item {\bf For $t=1,\ldots,T$:}
\begin{enumerate}
\item {\bf Sampling:}
\begin{enumerate}
 \item  Draw ${\bm \theta}_{t}^{(1)},...,{\bm \theta}_{t}^{(N)} \sim q({\bm \theta}|{\bm \mu}_t,{\bm \Lambda}_t)$.
\item  Assign to each sample the weights 
  \begin{eqnarray}
w_{t}^{(n)}=\frac{\pi_t({\bm \theta}_t^{(n)})}{q({\bm \theta}_t^{(n)}|{\bm \mu}_t,{\bm \Lambda}_t)}
, \qquad n=1,...,N.
\end{eqnarray}
\end{enumerate}
\item {\bf Current maximum estimations:} 
\begin{enumerate}
\item\label{StepMaxIter} Obtain ${\bm \theta}^{(t)}_{\texttt{max}} =\arg\max\limits_{n} \pi_t({\bm \theta}_t^{(n)})$,  and compute
$\widehat{\bf{r}}_t={\bf f}_r({\bm \theta}^{(t)}_{\texttt{max}})$.
\item \label{EsteSTEP_Sigma}
	Compute $\widehat{{\bm \Sigma}}_t=\dfrac{1}{R} \displaystyle\sum_{r=1}^{R} (\bm{y}_r - \widehat{\bf{r}}_t) (\bm{y}_r - \widehat{\bf{r}}_t)^{\top}$.
\end{enumerate}
\item {\bf Global maximum estimations:}
\begin{itemize}
\item\label{GlobCheck} If $\pi_t\big({\bm \theta}^{(t)}_{\texttt{max}}\big) > \pi_{\texttt{MAP}}$:
\begin{enumerate}
	\item $\widehat{{\bm \theta}}_{\texttt{MAP}}^{(t)}={\bm \theta}^{(t)}_{\texttt{max}} $, 
	\item $\widehat{{\bm \Sigma}}_{\texttt{ML}}^{(t)} = \widehat{{\bm \Sigma}}_t$,	
	 \item Update 
 $\pi_{\texttt{MAP}} = \pi_{t+1}\big(\widehat{{\bm \theta}}_{\texttt{MAP}}^{(t)}\big)$ (note that $\pi_{t+1}({\bm \theta})$ takes  into account the new $\widehat{{\bm \Sigma}}_{\texttt{ML}}^{(t)}$).
\end{enumerate}
\item Otherwise $\widehat{{\bm \theta}}_{\texttt{MAP}}^{(t)}=\widehat{{\bm \theta}}_{\texttt{MAP}}^{(t-1)}$, and  $\widehat{{\bm \Sigma}}_{\texttt{ML}}^{(t)} =\widehat{{\bm \Sigma}}_{\texttt{ML}}^{(t-1)}$.
\end{itemize}
\item {\bf Adaptation:} Set 
\begin{eqnarray} 
{\bm \mu}_{t+1}&=&\widehat{{\bm \theta}}^{(t)}_{\texttt{MAP}}, \label{muEqTable}\\
{\bm \Lambda}_{t+1}&=&\sum_{n=1}^N \bar{w}_t^{(n)} ({\bm \theta}_t^{(n)}-{\bar {\bm \theta}}_t)^{\top} ({\bm \theta}_t^{(n)}-{\bar {\bm \theta}}_t) + \delta_t {\bf I}_M, \label{SigmaEqTable}
\end{eqnarray}
where  $\bar{w}_t^{(n)} \frac{w_t^{(n)}}{\sum_{i=1}^N w_t^{(i)}}$ are the normalized weights, ${\bar {\bm \theta}}_t=\sum_{n=1}^N \bar{w}_t^{(n)} {\bm \theta}_t^{(n)}$ and $\delta_t >0$.
	\end{enumerate}
\item\label{OutputStep} {\bf Output:} Return the  final estimators $\widehat{{\bm \theta}}^{(T)}_{\texttt{MAP}}$, $\widehat{{\bm \Sigma}}^{(T)}_{\texttt{ML}}$, and all the weighted samples $\{{\bm \theta}_{t}^{(n)},\widetilde{w}_{t}^{(n)}\}$, for all $t$ and $n$, with the corrected weights
\begin{equation}\label{EQ_AQUIreweighting}
\widetilde{w}_{t}^{(n)}=w_{t}^{(n)} \frac{\pi_{T+1}({\bm \theta}_t^{(n)})}{\pi_t({\bm \theta}_t^{(n)})}.
\end{equation}
\end{enumerate} \\
\hline 
\end{tabular}
\end{table}

\bibliographystyle{unsrt}
\bibliography{bibliografia}





\end{document}